\documentclass[aps,pra,twocolumn,floatfix]{revtex4}
\usepackage{bm}
\usepackage{epsfig}

\begin{document}

\title{Quasars as bubbles of dark matter: evidence for axion and tachyon matter in
the Universe}
\author{Anatoly A. Svidzinsky}
\affiliation{Department of Physics, Institute for Quantum Studies,
Texas A\&M University, TX 77843-4242}
\date{\today }

\begin{abstract}
Growing amount of data show evidence for statistical and apparent physical
association between low-redshift galaxies and high-redshift quasi-stellar
objects, suggesting noncosmological origin of their redshift and failure of
classical quasar explanation. Here we find an analytical solution of
Einstein equations describing bubbles made from axions with periodic
interaction potential. Such particles are currently considered as one of the
leading dark matter candidate. The bubble interior has equal gravitational
potential and, hence, photons emitted from the interior possess identical
gravitational redshift. The redshift depends on the bubble mass and can have
any value between zero and infinity. Quantum pressure supports the bubble
against collapse and yields states stable on the scale of the Universe age.
Our results explain the observed quantization of quasar redshift and suggest
that intrinsically faint point-like quasars associated with nearby galaxies
(a few \% of known objects) are axionic bubbles with masses 10$^8$-10$%
^9M_{\odot }$ and radii 10$^3$-10$^4R_{\odot }$. They are born in active
galaxies and ejected into surrounding space. Properties of such quasars
unambiguously indicate presence of axion dark matter in the Universe and
yield the axion mass $m=0.4-3$ meV, which fits in the open axion mass window
constrained by astrophysical and cosmological arguments.

We also found that tachyons, another dark matter candidate, can form objects
with galactic size, negligible mass and any value of the gravitational
redshift. Such finding combined with quasar observations suggests that
bright quasars 3C 48, 3C 273 and 3C 279 are nuclei of forming nearby small
galaxies embedded into tachyonic clots and possess pure gravitational
redshift. If the bright quasars later evolve into small companion galaxies,
then their dark galactic halos, observed by rotation curves, are probably
remnants of the tachyon matter.
\end{abstract}

\maketitle

\section{Introduction}

Since their discovery in 1960's \cite{Schm63} quasars (QSOs) became one of
the most mysterious objects in the Universe. All observed QSOs share several
remarkable traits. First they are strong sources of radio waves and their
radiation spectrum is not a Planck spectrum of thermal radiation as in
ordinary stars. Secondly, all QSOs produce radiation with significant
redshift, some of them possess redshift $z>6$. Third, the radiation is
characterized by strong emission lines, while ordinary stars produce light
with strong absorption lines. Finally, all QSOs have small volumes by
galactic standards. One of the most puzzling features is that if the
redshift occurs due to large speed of QSOs relative to Earth, they must be
cosmological distances away and extremely luminous -- their luminosity
should lie between $10$ and $100$ of the total optical luminosity of the
brightest galaxies.

The present-day understanding of QSOs is based on the paradigm that redshift
of all QSOs has cosmological origin (the so called cosmological hypothesis).
Implied huge quasar's luminosity is the main consequence of the cosmological
hypothesis. Based on this assumption one can find some similarity between
QSOs and active galactic nuclei: both are intense energy producers, both
possess strong emission from radio to $X$-ray bands, sometimes both have
jets issuing from the central region. As a result, the present-day
conventional view regards QSOs and active galactic nuclei as having the same
nature.

The classical (generally accepted) picture of QSOs as active galactic nuclei
(AGN) is based on the hypothesis proposed by Sandage \cite{Sand73} that all
QSOs lie in the nuclei of galaxies. The weaker of them are the classical
Seyfert 1 nucleus, where the galaxy of stars surrounding the nucleus is
clearly seen. In the unified scheme the QSOs are high luminosity Seyfert 1
nuclei which contain a compact nuclear source ionizing a broad line region,
surrounded by an optically thick torus of dust. Depending on the orientation
of this torus with respect to the line of sight, the central object is seen
or hidden. When it is hidden, we see only the narrow, extended emission line
region; the galaxy is a Seyfert 2. When it is seen, we see both broad and
narrow emission lines. Most radio loud sources have a double lobe structure
which is powered by a relativistic jet. When the angle between the jet axis
and the line of sight is small the object appears as a blazar.

However, despite of great efforts, the nature of QSOs remains unclear so
far. The point is that the unifying scheme in its present form contradicts
to some observations which can not be ignored. Problems with the current
quasar's paradigm are discussed in many publications and here we mention the
main of them (for details see, e.g., \cite
{Arp87,Narl89,Burb92,Arp98,Kemb99,Burb01p}). One of the consequences of the
unified scheme is that host galaxies of all QSOs must be present. However,
observations with Hubble Space Telescope \cite{Bahc94,Hutc95} have shown
that for some QSOs there are no host galaxies, while for others there is
evidence for faint galaxies or irregular systems with the same redshift.
Another problem is distances to quasars. Some QSOs and the compact galaxies
they evolve into are resolved and many low redshift QSOs lie close to
galaxies with approximately the same redshifts \cite{Stoc78,Heck84,Yee87}.
This, as usually believed, suggests on cosmological distances to the
objects. On the other hand, there is a strong evidence that many low
redshift (nearby) galaxies and high redshift QSOs are physically associated
and, hence, these QSOs are no further away than the close galaxies and must
have redshifts noncosmological in origin (see, e.g., \cite
{Burb90,Burb92,Arp87,Burb96,Beni97,Arp98,Burb03,Arp04a,Arp04b,Gali04} and
references therein). Observations suggest that such quasars are ejected from
active galaxies or in the process of ejection from the galactic nucleus.

A systematic search for quasar-galaxy association was made by Burbidge et
al. \cite{Burb90}. They found over $500$ close pairs of QSOs and galaxies of
which only $28$ QSOs were associated with $42$ galaxies with approximately
equal redshifts (this is what the cosmological hypothesis predicts), while
for the reminder $92$\% objects the quasar redshifts considerably exceeded
the galaxy redshifts ($z_{\text{gal}}=0.0001-0.01$). Another evidence is the
observation of three QSOs with different large redshifts ($z=0.60,$ $1.40,$ $%
1.94$) near a spiral arm of the galaxy NGC 1073 by Arp and Sulentic \cite
{Arp87}. The galaxy NGC 1073 has much smaller redshift $z=0.004$ and
approximately $16$ Mpc away (for $H=75$km/s$\cdot $Mpc). There are dozens of
similar evidences, some of them are discussed in \cite
{Arp87,Arp98,Burb99,Arp02,Burb03,Arp04b,Gali04}. It is important to note
that there is a significant excess of galaxies only around flat spectrum
radio-loud QSOs (blazars), which constitute about $10\%$ of the known quasar
population. In contrary, there is a marginal defect of galaxies around
optically selected QSOs \cite{Beni97,Ferr97}.

One of the most fundamental quasar property is the redshift quantization.
Based on observations, Karlsson \cite{Karl90} has noted division of quasars
into two groups with different redshift properties and concluded the
following. If we select QSOs associated with most nearby (distance $%
d\lesssim 50-100$ Mpc), galaxies then their redshift is close to certain
values (quantized), as shown in Fig. \ref{hist} in next section. Meanwhile,
in QSO samples associated with distant, $d\gtrsim 50-100$ Mpc, galaxies no
periodicity in intrinsic redshift is observed. Such a division is supported
by later studies of QSOs associated with most nearby galaxies where the
quantization was confirmed \cite{Arp90,Burb01} and distant ($0.01<z_{\text{%
gal}}<0.3$) galaxies for which absence of any periodicity was claimed \cite
{Hawk02}. The property suggests existence of intrinsically faint (optical
luminosity $L=10^5-10^7L_{\odot }$) QSO subgroup with quantized
noncosmological redshift. Being intrinsically faint, such objects are not
detected from large distances which yields disappearance of redshift
quantization in distant QSO samples.

\begin{figure}[tbp]
\bigskip
\centerline{\epsfxsize=0.23\textwidth\epsfysize=0.21\textwidth
\epsfbox{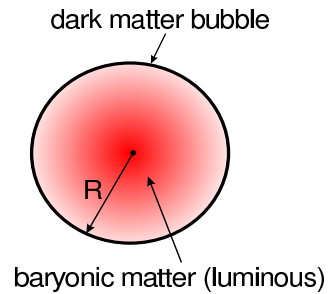}}
\caption{Structure of intrinsically faint quasars}
\label{fs}
\end{figure}

The observational evidences indicate that the total quasar population must
be divided into several groups having different nature but similar radiation
mechanisms. Quasars of the first group are intrinsically faint point-like
objects in the optical band and have structure unresolved in telescopes.
Such objects are born in nearby active galaxies and ejected into surrounding
space. Their redshift is \textit{quantized} and noncosmological (mostly
gravitational). Here we show that bubbles of dark matter, pictured in Fig.
\ref{fs}, with masses about $10^8-10^9M_{\odot }$ and radii $%
10^3-10^4R_{\odot }$ can explain the intrinsically faint point-like quasars.
The bubble is supported against collapse by quantum pressure and, in
principle, can have decay time larger then the age of the Universe.
Hypothetical axions, one of the leading dark matter candidate, fit well into
this picture and can account for the redshift quantization. Usual baryonic
matter falls into the bubble interior, heated by the release of the
gravitational energy and produce electromagnetic radiation. The bubble can
also harbor a black hole at the center which makes its structure similar to
AGN. The amount of baryonic matter trapped in the bubble we assume to be
small compared to the bubble mass. Bubble radiation spectrum must be similar
to AGN because the radiation mechanism is the same. Photons emitted anywhere
inside the bubble interior posses identical gravitational redshift and
freely propagate into surrounding space because dark matter is transparent
for electromagnetic waves.

Quasars of the second group are intrinsically bright objects with typical
optical luminosity $L=10^8-10^{10}L_{\odot }$ (a fraction of the objects can
be more luminous). Some of them are resolved in telescopes. Their redshift
possesses both the intrinsic and cosmological components, but the intrinsic
redshift is \textit{not quantized}. In Section 4 we show that not only small
on galactic scale objects can posses large gravitational redshift. Droplets
of tachyon dark matter with the mass much smaller then galactic mass can
produce any value of the gravitational redshift on kpc scales. Baryonic
matter embedded in such regions behaves as being possessed of an intrinsic
redshift. Tachyonic clots can explain the nature of intrinsically bright
QSOs which possibly have noncosmological redshift component but occupy large
space volumes $1-100$ pc in size. The classical model of quasars, as active
galactic nuclei with pure cosmological redshift, is a limiting case of
tachyonic quasars when the gravitational redshift component is negligible
compared to the cosmological contribution. Intrinsically bright quasars are
observed from substantially greater distances then the faint axionic bubbles
and, as a consequence, dominate in the known quasar population. We discuss
this issue in the last section.

Luminous active galactic nuclei with pure cosmological redshift belong to
the third group of quasars. However, as we mentioned, there is no sharp
distinction between such objects and the tachyonic quasars.

Being intrinsically faint, the axionic bubbles constitute only a few \% of
the observed QSOs population. Intrinsically bright QSOs are possibly a
heterogeneous family. Probably blazars, about 10\% of the known QSOs, have
the tachyonic nature and their redshift possesses substantial gravitational
component. First quasars discovered belong to this category. At the same
time, maybe most of the radio-quiet (steep radio spectra) QSOs (about 90\%
of the known QSOs population) are luminous active galactic nuclei at the
cosmological distances indicated by their redshifts. However, the last
question requires further analysis.

For the first time our theory was presented at the Fifth International
Heidelberg Conference on Dark Matter in Astro and Particle Physics (3-9
October, 2004) \cite{Svid04}; in this paper we provide all the details.

\section{Dark matter bubbles}

There is abundant evidence that the mass of the Universe is dominated by
dark matter of unknown form \cite{Sado99}. Particle dark matter, i.e., one
or more relic particle species from the big bang, is strongly suggested as
the dominant component of matter in the Universe \cite{Brad03}. For many
years weakly interacting massive particles (WIMPs), such as, for instance,
the lightest neutralino in the minimal supersymmetric standard model, were
the leading dark matter candidate. However, there is still no evidence in
favor of WIMP, either from bolometer experiments designed for direct
detection or from the observation of cosmic rays. Moreover, many recent
simulations of structure formation in the Universe suggested that any dark
matter component modelized as a gas of free particles, such as WIMPs, result
in cuspy density profiles at galactic centers, while observation of rotation
curves indicate a smooth core density \cite{Moor99}. Such a controversy
suggests that the dark matter halo probably consists of a Bose condensate of
light particles which behaves as a classical scalar field, coherent on the
scale of $10$ kpc. Self gravitating Bose condensate, governed by the
Klein-Gordon and Einstein equations, could account for the dark matter
distribution inside galaxies (see, e.g., \cite
{Arbe03,Mato01,Lope02,Fuch04,Mato03} and references therein). The field acts
as an effective cosmological constant (dark energy) before relaxing into a
condensate of nonrelativistic bosons \cite{Frie95,Vian99}. In order to
obtain the Bose condensate halo on the scale of $10$ kpc one should consider
ultralight noninteracting particles with masses $m\sim 10^{-23}$ eV or
heavier particles with self-coupling \cite{Hu00,Arbe03}. From the rotation
curve of the Andromeda Galaxy an estimate of the mean mass density of the
dark matter composing the $30$ kpc luminous core is $\rho \sim 2\times
10^{-24}$kg/m$^3$ \cite{Silv01}. Particle concentration $\rho /m$ determines
the temperature of Bose condensation transition $T_c$. The condition that $%
T_c$ is greater than the temperature of the cosmic microwave background
radiation imposes an upper limit on the particle mass $m<1$ eV. So, in order
to explain the dark matter halo by a Bose condensate of scalar bosons the
particle mass must be in the range $m=10^{-23}-1$ eV.

Here we study massive real scalar field $\varphi $ with periodic interaction
potential
\begin{equation}
\label{p1}V(\varphi )=V_0[1-\cos (\varphi /f)],
\end{equation}
where $V_0>0$. This potential is quite general and derived in quantum filed
theory in connection with pseudo Nambu-Goldstone bosons (PNGBs) \cite
{Frie95,Hill02,Hill88}. In all such models, the key ingredients are the
scales of global symmetry breaking $f$ and explicit symmetry breaking $%
(V_0)^{1/4}$. From the viewpoint of quantum field theory, PNGBs are the only
way to have naturally ultralow mass, spin 0 particles. One of the example of
a light hypothetical PNGB is the axion which possess extraordinarily feeble
couplings to matter and radiation and is well-motivated dark matter
candidate \cite{Brad03}. Axion arises from the Peccei-Quinn solution to the
strong CP problem \cite{Pecc77}. If the axion exists, astrophysical and
cosmological arguments constrain its mass $m$ to be in the range of $%
10^{-6}-3\times 10^{-3}$ eV and the global symmetry-breaking scale to lie in
a narrow window
\begin{equation}
\label{g1}f\approx \frac{m_\pi f_\pi }{2m}=2\times 10^9-6\times 10^{12}\text{%
GeV,}
\end{equation}
where $m_\pi =135$ MeV is the neutral pion mass and $f_\pi =93$ MeV its
decay constant \cite{Raff02,Brad03}. Axions in this mass range could provide
much or all of the cold dark matter in the Universe. Interaction of axions
with QCD instantons generates the axion mass and cosine interaction
potential \cite{Kim87}. Another example of ultralight PNGB is a hypothetical
scalar field that arises when the global symmetry is broken on a Planck
scale $f\sim 10^{18}$ GeV and explicit breaking scale is comparable to light
neutrino mass $(V_0)^{1/4}\sim 10^{-3}$ eV. Such a field, which acquires a
mass $m\sim \sqrt{V_0}/f\sim 10^{-24}$ eV, would currently dominate the
energy density of the Universe \cite{Frie95}.

A self-gravitating real scalar field $\varphi $ in general relativity is
described by the action \cite{Seid90} (in this section we use natural units
for which $\hbar =c=1$)
\begin{equation}
\label{a1}I=\int d^4x\sqrt{-J}\left\{ \frac 1{16\pi G}R_i-\left[ \frac
12g^{\mu \nu }\partial _\mu \varphi \partial _\nu \varphi +V(\varphi
)\right] \right\} ,
\end{equation}
where $G$ is the gravitational constant, $J$ is the determinant of the
metric tensor $g_{\mu \nu }$ and $R_i$ is the Riemann curvature. We consider
spherically symmetric system, so the metric can be written in the form

\begin{equation}
\label{b1}ds^2=-N^2dt^2+g^2dr^2+r^2d\Omega ^2,
\end{equation}
where $g$, the radial metric, and $N$, the lapse, are functions of $t$ and $%
r $ with $r$ being the circumferential radius. We introduce dimensionless
coordinates and define the unit of distance, time and $\varphi $ as
\begin{equation}
\label{a2}r_0=\frac \hbar {mc},\quad t_0=\frac \hbar {mc^2},\quad \varphi
_0=\frac 1{\sqrt{4\pi G}},
\end{equation}
where $c$ is the speed of light, $m=\sqrt{V_0}/f$ is the mass of the scalar
particle. In dimensionless units the coupled Klein-Gordon and Einstein
equations describing dynamics of the field $\varphi $ and the metric are
\cite{Seid90}
$$
-\frac{\ddot \varphi }{N^2}+\frac{\dot N\dot \varphi }{N^3}+\frac{\varphi
^{\prime }}{g^2}\left( \frac{g^2+1}r-2rg^2V\right) +\frac{\varphi ^{\prime
\prime }}{g^2}-\quad
$$
\begin{equation}
\label{b2}\frac{r\dot \varphi ^2\varphi ^{\prime }}{N^2}-\frac{\partial V}{%
\partial \varphi }=0,
\end{equation}
\begin{equation}
\label{b3}N^{\prime }=\frac N2\left[ \frac{g^2-1}r+r\left( \varphi ^{\prime
2}-2g^2V+\frac{g^2\dot \varphi ^2}{N^2}\right) \right] ,
\end{equation}
\begin{equation}
\label{b4}g^{\prime }=\frac g2\left[ \frac{1-g^2}r+r\left( \varphi ^{\prime
2}+2g^2V+\frac{g^2\dot \varphi ^2}{N^2}\right) \right] ,
\end{equation}
where an overdot denotes $\partial /\partial t$, a prime denotes $\partial
/\partial r$,
\begin{equation}
\label{a3}V=\frac 1{\alpha ^2}[1-\cos (\alpha \varphi )],\quad \alpha =\frac
1{\sqrt{4\pi G}f}=\frac{m_{\text{pl}}}{\sqrt{4\pi }f}
\end{equation}
is the dimensionless potential and the coupling parameter respectively, $m_{%
\text{pl}}=\sqrt{\hbar c/G}=1.2\times 10^{19}$ GeV is the Planck mass. The
interaction potential $V$ has degenerate minima at $\varphi =2\pi n/\alpha $%
, where $n$ is an integer number. It is tempting to search for
time-independent solutions of Eqs. (\ref{b2})-(\ref{b4}). However, the
pseudovirial theorem of Rosen \cite{Rose66} implies that no such solution is
possible in the Newtonian limit, and in the strong-field case it has been
shown numerically that no nonsingular solution exists \cite{Koda78}. Here we
show that in the limit of strong nonlinearity ($\alpha \gg 1$) an
approximate time-independent solution does exist and it describes a
spherical bubble with surface width much smaller then its radius $R$. The
bubble surface is an interface between two degenerate vacuum states with $%
\varphi =2\pi n/\alpha $ ($r<R$) and $\varphi =0$ ($r>R$).

We look for static solutions, then Eqs. (\ref{b2})-(\ref{b4}) reduce to
\begin{equation}
\label{b5}\frac{\varphi ^{\prime }}{g^2}\left( \frac{g^2+1}r-2rg^2V\right) +%
\frac{\varphi ^{\prime \prime }}{g^2}-\frac{\partial V}{\partial \varphi }%
=0,
\end{equation}
\begin{equation}
\label{b6}N^{\prime }=\frac N2\left[ \frac{g^2-1}r+r\left( \varphi ^{\prime
2}-2g^2V\right) \right] ,
\end{equation}
\begin{equation}
\label{b7}g^{\prime }=\frac g2\left[ \frac{1-g^2}r+r\left( \varphi ^{\prime
2}+2g^2V\right) \right] ,
\end{equation}
with the following boundary conditions

$$
g(0)=g(\infty )=N(\infty )=1,\quad
$$
$$
g^{\prime }(0)=N^{\prime }(0)=\varphi ^{\prime }(0)=0,\quad V(\varphi
(\infty ))=0.
$$
Outside the bubble $\varphi =0$ and Eqs. (\ref{b5})-(\ref{b7}) lead to the
Schwazschild solution for a spherically symmetric problem:
\begin{equation}
\label{sch1}g^2=\frac 1{1-2M/r},\quad N^2=1-\frac{2M}r,
\end{equation}
where $M$ is the dimensionless bubble's mass in units of $m_{\text{pl}}^2/m$.

Eqs. (\ref{b5})-(\ref{b7}) can be obtained as an extremum condition of the
energy functional%
$$
E[N,g,\varphi ]=\int_0^\infty dr\frac Ng\left\{ r^2\left( \frac{\varphi
^{\prime 2}}2+g^2V\right) -\frac 12(g-1)^2\right. +
$$
\begin{equation}
\label{e1}\left. r(g-1)\frac{N^{\prime }}N\right\} .
\end{equation}
Variation of this functional with respect to $\varphi $ gives the
Klein-Gordon equation (\ref{b5}), while variation with respect to the metric
functions $g$ and $N$ yields the Einstein equations (\ref{b6}), (\ref{b7})
respectively. Using Eqs. (\ref{b7}), (\ref{e1}) the total bubble energy
reduces to
\begin{equation}
\label{e2}E=\int_0^\infty dr\left[ r\left( 1-\frac 1g\right) N\right]
^{\prime }=\lim _{r\rightarrow \infty }r\left( 1-\frac 1g\right) =M,
\end{equation}
here we applied the Schwazschild solution (\ref{sch1}) valid outside the
bubble. Eq. (\ref{e2}) demonstrates Einstein equivalence principle between
the mass and the energy. Eq. (\ref{b7}) leads also to another expression for
the bubble mass:
\begin{equation}
\label{e3}M=\int_0^\infty dr\left[ r\left( 1-\frac 1{g^2}\right) \right]
^{\prime }=\int_0^\infty drr^2\left( \frac{\varphi ^{\prime 2}}{g^2}%
+2V\right) ,
\end{equation}
which shows that the scalar field gradient and the potential $V$ are sources
of the bubble energy.

Let us assume that $\alpha \gg 1$ and the radius of the bubble is $R\gg
\alpha $. Then inside the bubble, including its surface region, one can omit
terms with $1/r$ in Eqs. (\ref{b5})-(\ref{b7}) and take $r\approx R$
(thickness of the bubble surface is much smaller than its radius), we obtain
\begin{equation}
\label{b8}-2R\varphi ^{\prime }V+\frac{\varphi ^{\prime \prime }}{g^2}-\frac{%
\partial V}{\partial \varphi }=0,
\end{equation}
\begin{equation}
\label{b9}N^{\prime }=\frac{NR}2\left( \varphi ^{\prime 2}-2g^2V\right) ,
\end{equation}
\begin{equation}
\label{b10}g^{\prime }=\frac{gR}2\left( \varphi ^{\prime 2}+2g^2V\right) .
\end{equation}
Eqs. (\ref{b8})-(\ref{b10}) can be solved analytically. Their first integral
is
\begin{equation}
\label{b11}N=const,
\end{equation}
\begin{equation}
\label{b12}\varphi ^{\prime 2}=2g^2V,
\end{equation}
\begin{equation}
\label{b13}g^{\prime }=Rg\varphi ^{\prime 2}.
\end{equation}
Further, Eqs. (\ref{b13}) and (\ref{b12}) yield
\begin{equation}
\label{e4}\frac 1g=1-R\int_\varphi ^{\varphi (0)}\sqrt{2V}d\varphi .
\end{equation}
We assume that $\varphi (0)=2\pi n/\alpha $, where $n=1,2,3,\ldots $ is the
number of kinks at the bubble surface, and $\varphi (r)$ monotonically
decreases with $r$. Outside the bubble $\varphi =0$. Substitute (\ref{e4})
into (\ref{b12}) leads to
\begin{equation}
\label{e4ff}\varphi ^{\prime }=-\frac{\sqrt{2V}}{1-R\int_\varphi ^{\varphi
(0)}\sqrt{2V}d\varphi }.
\end{equation}
Further integration can be made for a particular choice of the scalar field
potential. For $V(\varphi )$ given by Eq. (\ref{a3}) the final solution is%
$$
\frac{4R}{\alpha ^2}\ln |\sin (\alpha \varphi /2)|+\left[ 1-\frac{4R}{\alpha
^2}(2m-1)\right] \text{arctanh}[\cos (\alpha \varphi /2)]
$$
$$
=\text{sign}[\sin (\alpha \varphi /2)](r-R_m),
$$
\begin{equation}
\label{b14}\varphi \in [2\pi (n-m+1)/\alpha ,2\pi (n-m)/\alpha ],
\end{equation}
where $R_m$ is a position of the $m$th kink, $m=1,2,\ldots ,n$. When the
coordinate $r$ passes through the point $R_m$ the scalar field $\varphi (r)$
changes from $2\pi (n-m+1)/\alpha $ to $2\pi (n-m)/\alpha $ (see Fig. 2a).
Eq. (\ref{e4}) yields the following expression for $g$ as a function of $%
\varphi $ inside the bubble:
\begin{equation}
\label{b15}\frac 1g=1-\frac{4R}{\alpha ^2}[2m-1+\cos (\alpha \varphi /2)].
\end{equation}
Outside the bubble $\varphi =0$, $m=n$ and $1/g=1-8nR/\alpha ^2$. The
solution is valid if $1/g>0$, that is $R<R_{\max }=\alpha ^2/8n$. Match of
the inner solution (\ref{b15}) with the Schwazschild solution (\ref{sch1})
outside the bubble determines the bubble mass:
\begin{equation}
\label{b16}M=\frac{8nR^2}{\alpha ^2}-\frac{32n^2R^3}{\alpha ^4}.
\end{equation}
The mass depends on the bubble radius $R$ and changes between zero for $R=0$
and a maximum value $M_{\max }=\alpha ^2/16n$ for $R=R_{\max }$.

The mass-radius relation (\ref{b16}) was derived for the cosine interaction
potential (\ref{a3}). However, using Eq. (\ref{e4}), one can obtain similar
relation for any periodic potential $V(\varphi )$%
\begin{equation}
\label{b16g}M=4\pi nuR^2-8\pi ^2n^2u^2R^3,
\end{equation}
where $u$ is the surface energy density (the energy per unit area)
determined by an integral over one potential period
\begin{equation}
\label{u}u=\frac 1{4\pi }\int \sqrt{2V}d\varphi .
\end{equation}
For the cosine potential (\ref{a3}) $u=2/\pi \alpha ^2$.

Redshift of the bubble interior $z=1/N-1$ can be found by matching the inner
solution $N=const$ with the Schwazschild solution (\ref{sch1}) outside the
bubble. As a result, we obtain that everywhere inside the bubble the
redshift is the same and given by
\begin{equation}
\label{b17}z=\frac 1{\sqrt{1-2M/R}}-1=\frac 1{1-4\pi nuR}-1.
\end{equation}
The internal redshift monotonically increases from zero to infinity when the
bubble radius $R$ changes from zero to $R_{\max }$. Fig. 2b shows redshift
of space as a function of the distance $r$ to the bubble center. The
redshift is constant in the bubble interior and monotonically decreases
outside the bubble.

\begin{figure}[tbp]
\bigskip
\centerline{\epsfxsize=0.45\textwidth\epsfysize=0.33\textwidth
\epsfbox{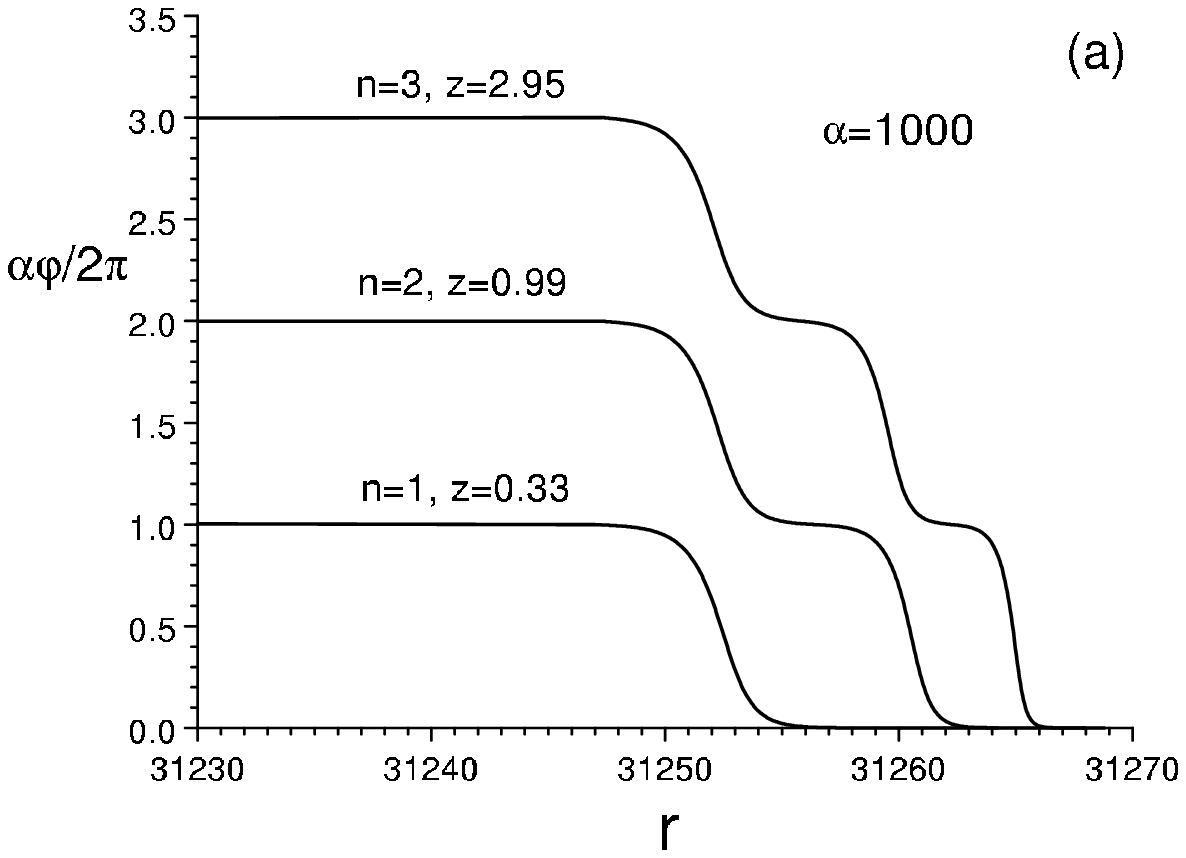}}
\end{figure}

\begin{figure}[tbp]
\bigskip
\centerline{\epsfxsize=0.45\textwidth\epsfysize=0.33\textwidth
\epsfbox{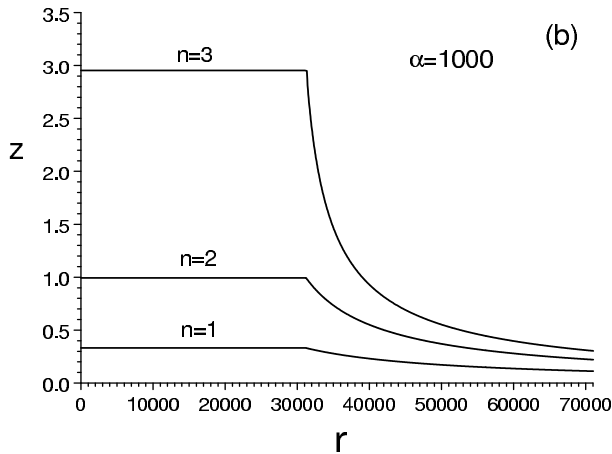}} 
\caption{(a) Scalar field $\protect\varphi $ as a function of distance $r$
to the bubble center for bubbles with equal radius and different quantum
numbers $n=1$, $2$, $3$. The unit of length is $\hbar/mc$. Note, we plot the
field $\protect\varphi $ only in the vicinity of the bubble surface where it
undergoes variation. (b) Redshift $z$ of space as a function of distance $r$
to the bubble center for bubbles shown in Fig. 2a. The redshift is constant
in the bubble interior. }
\end{figure}

\begin{figure}[tbp]
\bigskip
\centerline{\epsfxsize=0.45\textwidth\epsfysize=0.34\textwidth
\epsfbox{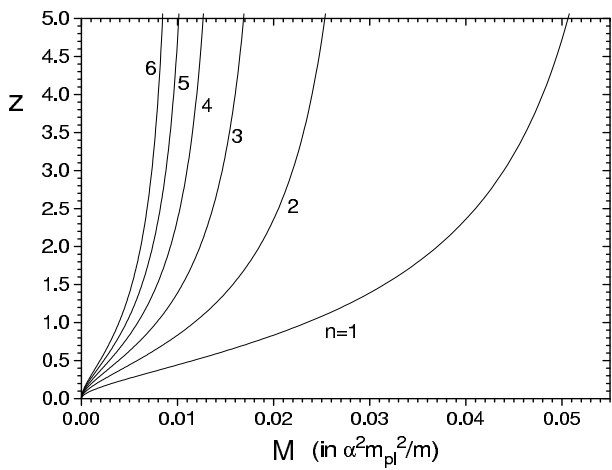}}
\par
\caption{Redshift of a bubble as a function of mass for different ``quantum"
numbers $n=$1, 2, 3, 4, 5, 6. }
\label{rn}
\end{figure}

In Fig. \ref{rn} we plot the bubble gravitational redshift as a function of
its mass for different kink number $n$. The redshift monotonically increases
with the mass and goes to infinity when the mass approaches the maximum
value $M_{\max }=\alpha ^2/16n$.

\subsection{Redshift quantization}

Let us make rescaling $M\rightarrow M/4\pi u$, $R\rightarrow R/4\pi u$, then
Eqs. (\ref{b16g}), (\ref{b17}) become
\begin{equation}
\label{b18}M=nR^2-n^2R^3/2,
\end{equation}
\begin{equation}
\label{b19}z=\frac 1{1-nR}-1.
\end{equation}
For a given $M$ the redshift depends on the number of kinks $n$, which
implies that the redshift is quantized.

In early QSO samples involving about 600 radio-emitting QSOs associated with
bright nearby spiral galaxies ($z_{\text{gal}}<0.007$), Karlsson showed that
the redshift distribution has a periodicity $\log (1+z_{n+1})-\log
(1+z_n)=0.089$, where $n=0,1,2,$ $\ldots $ and $z_0=0.061$ \cite
{Karl90,Karl71}. It has been later confirmed by other groups \cite
{Barn76,Arp90}. In a recent paper, Burbidge and Napier \cite{Burb01} tested
for the occurrence of this periodicity in new QSO samples and found it to be
present at a high confidence level. The peaks were found at $z\approx 0.30$,
$0.60$, $0.96$, $1.41$ and $1.96$ in agreement with Karlsson's empirical
formula. The formula also includes the peak at $z_0=0.061$, however, this
peak does not occur for quasars, but for morphologically related objects.

The redshift periodicity is observed only in QSO samples satisfying certain
selection criteria, in particular, the galaxies which are assumed to be
paired to the QSOs must be \textit{most nearby} spirals \cite{Karl90,Napi03}%
. This implies that redshift quantization is a property of intrinsically
faint QSOs which are not detected from large, $\gtrsim 50-100$ Mpc,
distances. The observed peaks are narrow, implying that the spread in
cosmological redshift and Doppler redshift due to random motion must both be
small. If they were not, the intrinsic peaks would be washed out, as is
easily seen from the combined redshift formula
$$
1+z=(1+z_g)(1+z_c)(1+z_d),
$$
where $z$, $z_g$, $z_c$ and $z_d$ denote the total, gravitational,
cosmological and Doppler redshift respectively. The morphology of companion
galaxies is probably also important \cite{Napi03}.

If we consider a sample of quasars born in the same type of galaxies it is
naturally to expect that such objects would have approximately equal masses
because their formation mechanism must be similar. Such phenomenon is well
known for type Ia supernovae or neutron stars: practically all measured
neutron star masses cluster around the value of $1.4M_{\odot }$ with only a
few percent deviation \cite{Glen00}. If we assume that dark matter bubbles
are born with equal masses then their redshift must be quantized. For a
given $M$ Eqs. (\ref{b18}), (\ref{b19}) have a set of solutions for $R$ and $%
z$ corresponding to different values of the ``quantum'' number $n$. For $%
M=0.0601$ (in dimension units $M=0.00752${\small $\alpha ^2m_{\text{pl}}^2/m$%
)} Eqs. (\ref{b18}), (\ref{b19}) have solutions for $n=1,$ $2,$ $\ldots ,$ $%
8 $, they are given in Table 1.


\begin{table}[tbp]
\centerline{
\begin{tabular}{|c|c|c|}
\hline
n & R, in $\alpha^2\hbar/mc$ & z \\
\hline
1 & 0.0329 & 0.357  \\
2 & 0.0241 & 0.629  \\
3 & 0.0204 & 0.96  \\
4 & 0.0182 & 1.40  \\
5 & 0.0168 & 2.06  \\
6 & 0.0159 & 3.24  \\
7 & 0.0153 & 6.11  \\
8 & 0.0151 & 26.6  \\
\hline
\end{tabular}
}
\caption{Redshift of the bubble interior $z$ and its radius $R$ for $M=0.00752\protect\alpha ^2m_{\text{pl}}^2/m$ and different kink numbers $n$.}
\end{table}

\begin{figure}[tbp]
\bigskip
\centerline{\epsfxsize=0.45\textwidth\epsfysize=0.37\textwidth
\epsfbox{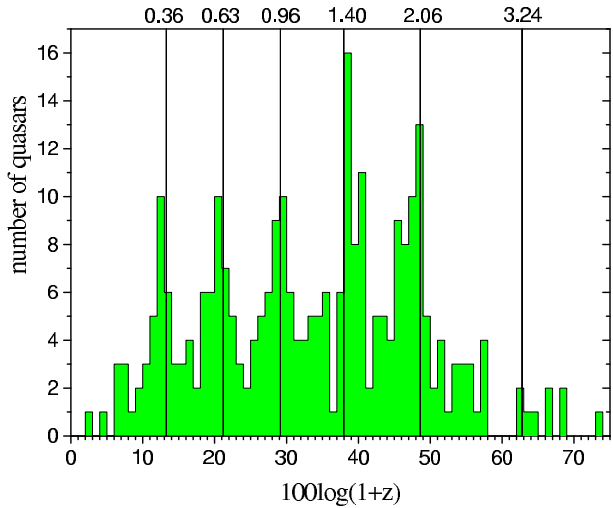}}
\par
\caption{Histogram of the redshift distribution of QSOs close to bright
active spiral galaxies or multiple QSOs with small angular separation from
Ref. \protect\cite{Napi03}. The solid lines represent position of the peaks
from Table 1. }
\label{hist}
\end{figure}

In Fig. \ref{hist} we plot the most recent histogram of the redshift
distribution from Ref. \cite{Napi03} in which five peaks are clearly seen.
The solid lines show the redshifts from our Table 1, they match well the
observed peaks. The agreement is remarkable because the theory has only one
free parameter, the bubble mass $M$. Such coincidence strongly suggests that
some fraction of quasar population is dark matter bubbles composed of scalar
particles with periodic interaction potential. One should mention an
alternative possibility of quasar evolution. Bubbles can be originally born
with the same mass and number of kinks $n=5$ that corresponds to the $5$th
peak. During evolution the kinks tunnel to the bubble center and quasars
sequentially decay into states with smaller $n$ but the same mass.

If a bubble is made of axions with $m=10^{-4}-3\times 10^{-3}$ eV and $%
f\approx f_\pi m_\pi /2m=2\times 10^9-6\times 10^{10}$ GeV, then, according
to Eq. (\ref{a3}), $\alpha =5.6\times 10^7-1.7\times 10^9$. Then a bubble
with the internal redshift $z=0.36$ and $n=1$ would have the mass $%
M=0.00752\alpha ^2m_{\text{pl}}^2/m=3\times 10^7-10^9M_{\odot }$, surface
width $\sim \hbar /mc=0.07-2$mm, surface mass density $4\times
10^{12}-10^{14}$kg/m$^2$ and the radius $R=0.0329\alpha ^2\hbar /mc=3\times
10^2-10^4R_{\odot }$. Such radius range agrees with the size of the emission
region expected for the intrinsically faint point-like quasars associated
with nearby galaxies. Indeed, analysis of the emission lines intensity of
the bright quasar 3C 48 shows that if it was located a distance of $1000$
Mpc then the size of the broad emission-line region should be about $0.1$ pc
\cite{Thua79}. The quasars we consider as candidates into dark matter
bubbles have the apparent visual magnitude 18 - 21 \cite{Arp04b,Burb03}
which yields the brightness 1-2 orders smaller that 16th magnitude quasar 3C
48. If they are located $1-10$ Mpc away and radiating gas has the same
parameters as for 3C 48 their size must be of the order of $10^4R_{\odot }$
(we assume that the brightness scales as $R^3/d^2$, where $d$ is the
distance to the object).

An empirical relation between the size of the broad-line region $R$ and
luminosity $L$ of Seyfert 1 galaxies, $R\propto L^{1/2}$ \cite{Wang03}
yields a similar answer. Indeed, for Seyfert 1 galaxies $R\sim 10-100$ light
days. The luminosity of the point-like quasars is 5-6 orders smaller, which
for the quasars leads to $R\sim 10^4R_{\odot }$.

\section{Stability}

\begin{figure}[tbp]
\bigskip
\centerline{\epsfxsize=0.47\textwidth\epsfysize=0.33\textwidth
\epsfbox{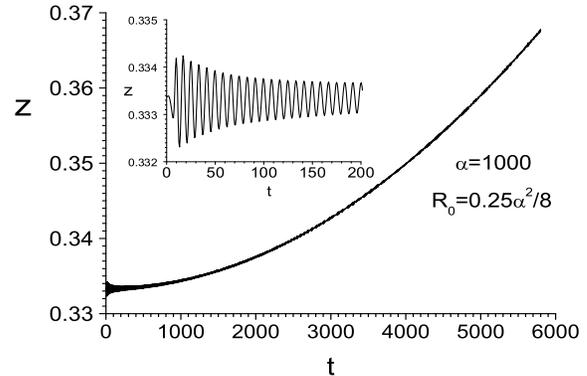}}
\par
\caption{Redshift of the bubble interior $z$ as a function of time $t$
during collapse of initialy static one kink bubble with the
radius $R_0=0.25\protect\alpha^2/8$ and $\protect\alpha=1000$.
The unit of time is $\hbar/mc^2$. }
\label{ztm}
\end{figure}

\begin{figure}[tbp]
\bigskip
\centerline{\epsfxsize=0.45\textwidth\epsfysize=0.3\textwidth
\epsfbox{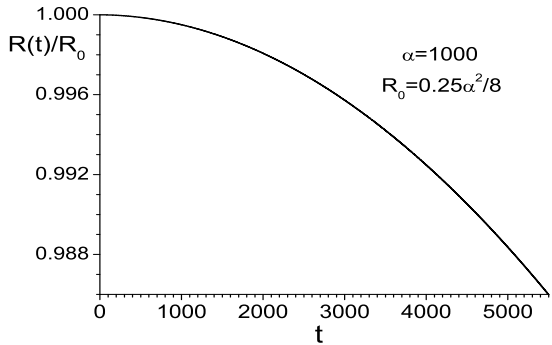}}
\par
\caption{Radius of the bubble $R$ as a function of time $t$ during collapse
of initialy static bubble with the radius $R_0=0.25\protect\alpha^2/8$ and $\protect\alpha=1000$. The unit of time is $\hbar/mc^2$. }
\label{rtm}
\end{figure}

The obtained analytical solution of the stationary equations (\ref{b5})-(\ref
{b7}) is approximate because the terms of the order of $1/\alpha ^2$ were
omitted. For $\alpha \gg 1$ such terms are very small, however, they play a
crucial role in the bubble stability. Our numerical calculations show that
the complete system of stationary equations (\ref{b5})-(\ref{b7}) has no
solution which indicates a possible instability. To study the unstable mode
we solve numerically the evolution equations for the scalar field (\ref{b2}%
)-(\ref{b4}) with the initial condition given by the approximate analytical
solution (\ref{b14}), (\ref{b15}); the result is shown in Figs. \ref{ztm}
and \ref{rtm}. Under the influence of surface tension and gravitational
attraction the initially static bubble starts to collapse very slowly with
the acceleration $a=\ddot R\sim -c^2/R$ (the corresponding unstable mode has
an imaginary frequency $\omega \sim ic/R$). The contraction is accompanied
by radiation of scalar particles from the bubble surface which propagate
into surrounding space. One can describe the instability as a tendency to
reduce the surface area of a hollow bubble with the surface tension $\sigma
=u$.

The dynamics of relativistic bubbles has been investigated in connection
with the bag model of hadrons and phase transitions in the early Universe
\cite{Auri84,Blau87,Auri87,Bere87,Auri89}. The study has mostly focused on
the thin-wall approximation where the bubble surface is specified by
effective parameters such as the surface energy density $u$ and the surface
tension $\sigma $. The approximation omits the particle emission and
describes the bubble dynamics by a single differential equation for the
radius $R$. In particular, when $\dot R=0$ the acceleration is given by \cite
{Blau87}

\begin{equation}
\label{bL1}u\ddot R=-\frac{2N^3\sigma }R-\frac{N^2Mu}{(1+N)R^2},
\end{equation}
where $N=\sqrt{1-2M/R}$. The first term on the right-hand side of Eq. (\ref
{bL1}) is the Laplace pressure, while the second term is the pressure caused
by gravitational attraction. At zero temperature $u=\sigma $. Our numerical
result for the surface acceleration (valid at $T=0$) agrees well
quantitatively with Eq. (\ref{bL1}) obtained in the thin-wall approximation.
This indicates that energy dissipation due to emission of scalar particles
is negligibly small and does not affect the bubble contraction.

\begin{figure}[tbp]
\bigskip
\includegraphics[angle=0,width=8cm]{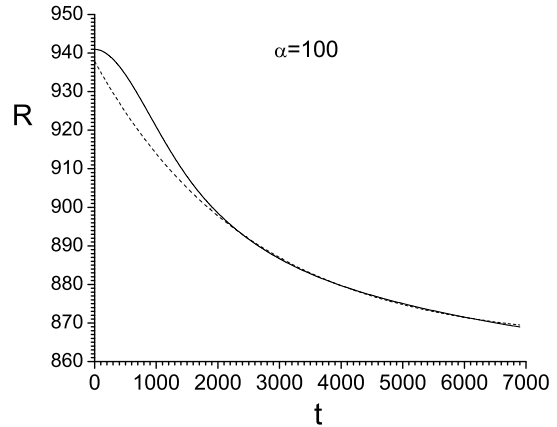}

\caption{Radius of the bubble $R$ as a function of time $t$ during collapse
of initialy static bubble with the
radius $R_0=941$, initial redshift $z_0=3.07$ and $\protect\alpha=100$
(solid curve). Dashed line shows exponentially decaying fit
$R(t)=865+73\exp(-t/2523)$.
The unit of length is $\hbar/mc$, while the unit of time is $\hbar/mc^2$. }
\label{exp}
\end{figure}

When the bubble size approaches the gravitational radius $R_g=2M$ a distant
observer sees that shrinking bubble starts to slow down and the shrinking
speed decreases to zero. Fig. \ref{exp} shows solution of the evolution
equations (\ref{b2})-(\ref{b4}) in the regime when such a behavior is
achieved. The bubble radius never crosses the event horizon; in the
reference frame of the distant observer the radius $R(t)$ only
asymptotically approaches $R_g$. At large $t$ the bubble radius $R(t)$ is
well approximated by an exponentially decaying function shown by dashed line
in Fig. \ref{exp}. The effect is caused by time dilation. Motion of a
particle falling into a black hole with a gravitational radius $R_g$ is a
classical example of such effect. A distant observer sees that the particle
asymptotically approaches $R_g$, but never crosses it; the particle
distance to the center is given by \cite{Land88}
\begin{equation}
\label{rtg}R(t)=R_g+\text{const}\times \exp (-ct/R_g).
\end{equation}
Radius of the collapsing bubble obeys similar asymptotic behavior; only the
coefficient in the exponent differs by a factor of the order of one.

One should mention that if we neglect gravity the bubble would collapse into
the origin and disintegrate on a time scale of a few times of the light
crossing time $R/c$ \cite{Widr89}. However, if gravity is included this
leads to appearance of the event horizon which prevents the bubble collapse
into the origin. The bubble life-time in the presence of gravity becomes
much larger than $R/c$. The large value of the bubble life time is a
consequence of time dilation produced by gravitation field. In the proper
coordinate frame the bubble decays on a short time scale $\sim R/c$.
However, in the reference frame of a distant observer (which is relevant to
quasar problem) the life time is much longer.

For the thin-wall bubble the total mass-radius relation is given by \cite
{Blau87,Auri91}
\begin{equation}
\label{bL2}M=\frac{4\pi uR^2}{\sqrt{1-(dR/d\tau )^2}}-8\pi ^2u^2R^3,
\end{equation}
where $\tau $ is the interior coordinate time ($d\tau =Ndt$). When velocity
of the bubble surface $dR/d\tau =0$ the expression (\ref{bL2}) reduces to
our mass-radius relation (\ref{b16g}) for $n=1$.

\subsection{Quantum stabilization}

A classical bubble shrinks towards its gravitation radius $R_g$ and at $t\gg
R_g/c$ behaves as a black hole. Such black hole-like objects can not explain
the nature of intrinsically faint quasars. Here we show however that quantum
effects suppress the collapse and result in appearance of long-lived bubbles
(not black holes), stable on the scale of the Universe age. So far we
considered the scalar field as classical and Eq. (\ref{bL2}) defines the
classical equation of motion of the bubble radius. However, quantum
corrections can be crucial for the bubble stability since the later is
determined by small terms in Eqs. (\ref{b5})-(\ref{b7}). It is known that
soliton bubbles which are unstable in the classical model can be stabilized
by quantum corrections \cite{Balt91,Ague95}. From the first glance it is
somewhat counter-intuitive that for objects $10^4R_{\odot }$ in size quantum
description is appropriate. However, interaction of dark matter with
incoherent environment is extremely weak. Therefore, a coherent state of
dark matter can live for a long time, which indicates that the quantum
description is probably adequate.

To include quantum effects it has been suggested that the expression (\ref
{bL2}) for the mass-energy be interpreted as the canonical hamiltonian of
the bubble at the quantum level \cite{Bere88,Auri90,Auri91}. The bubble wave
function $\Psi (R,\tau )$ satisfies the following quantum mechanical
equation in 1+1 dimensions ($\hbar =1$) \cite{Auri91}:
\begin{equation}
\label{bL3}\left[ \left( -i\frac \partial {\partial \tau }+8\pi
^2u^2R^3\right) ^2+\frac{\partial ^2}{\partial R^2}-16\pi ^2u^2R^4\right]
\Psi (R,\tau )=0.
\end{equation}
This equation is interpreted as a one-dimensional wave equation for a
relativistic particle moving along the semi-axis $R>0$ with the boundary
condition $\Psi (R=0)=0$. Separating variables in Eq. (\ref{bL3}) leads to
the stationary one-dimensional wave equation \cite{stat}
\begin{equation}
\label{bL4}\left[ \left( E+8\pi ^2u^2R^3\right) ^2+\frac{\partial ^2}{%
\partial R^2}-16\pi ^2u^2R^4\right] \Psi (R)=0.
\end{equation}
This equation possesses stationary solutions that are not possible in the
classical model and the energy spectrum of the bubble turns out to be
discrete \cite{Coll76,Auri90}. Such stationary quantum bubbles can explain the
nature of intrinsically faint quasars.
At the quantum level the collapse is
prevented by the uncertainty principle which forbids the exact localization
of an extended object in space. The uncertainty principle yields an
effective quantum pressure which balances the surface tension and
gravitational attraction producing stationary configurations. The discrete
energy spectrum can be obtained from the quasiclassical quantization
condition
\begin{equation}
\label{bL5}\int_0^{R_0}p(R)dR+\pi /4=(k+1)\pi ,
\end{equation}
where momentum $p(R)$ is given by
\begin{equation}
\label{bL51}p(R)=\sqrt{\left( E+8\pi ^2u^2R^3\right) ^2-16\pi ^2u^2R^4},
\end{equation}
$k=0,$ $1$, $2$, ... is the level number and $R_0$ is the classical turning
point at which $p(R)=0$. The turning point is determined by the classical
mass-radius relation (\ref{b16g}). In the limit of weak gravity, $uR\ll 1$,
Eq. (\ref{bL5}) yields the following expression for the energy levels \cite
{Auri90}
\begin{equation}
\label{bL6}E=3.59u^{1/3}(k+3/4)^{2/3}.
\end{equation}
Bubbles of non-negligible gravitational redshift correspond to highly
excited stationary states with $k\sim (\alpha /10)^4\gg 1$. For such states
the spacing between consecutive levels is small compared to the energy and
the level distribution can be treated as quasi-continuous.

Let us now estimate the decay rate of the quantum bubble. Partially the
picture is analogous to the hydrogen atom where electron moves in the field
of nucleus. If initially the electron is localized in a small region at a
large distance $R$ from the nucleus then its further motion would be similar
to the motion of a classical point particle. The originally resting electron
will start to move radially and fall down to the nucleus during the time $t_{%
\text{c}}\sim \sqrt{R/a}$, where $a$ is the initial acceleration. However,
if initially the electron is in a highly excited but stationary state its
wave function is delocalized. The uncertainty principle pressure supports
such electron cloud from collapse. The decay of the state occurs by means of
consecutive transitions between stationary states with smaller principal
quantum numbers and energy loss by photon emission. Such decay time $t$ has
nothing to do with the time $t_{\text{c}}\,$of the classical radial fall of
the electron on the nucleus. Nevertheless, $t$ can be estimated from
classical equations as the time of energy loss by a point electron
performing radial oscillations around the nucleus and emitting
electromagnetic waves (even if in the quantum picture there are no such
oscillations). Such a rule is a manifestation of the Bohr correspondence
principle \cite{Bere89}. This yields $t\sim t_{\text{c}}^4/(R/c)^3\sim
10^{-8}$s which agrees with the quantum mechanical answer for the decay rate
of the hydrogen atom excited states, while $t_{\text{c}}\sim 10^{-16}$s.

The decay of an excited stationary state of the quantum bubble occurs by
means of scalar particle emission. We estimate the decay rate using the
classical picture as the time of energy loss by the classical bubble with
the radius $R(t)$. We are interested in the bubble life-time measured by an
outside distant observer. Any correct quantum description cannot avoid the
Bohr correspondence principle. So, if the classical bubble treatment,
mentioned above, yields very long life-time ($\gg R/c$), the quantum bubble
must also be very long-lived.

The outside observer cannot witness crossing of the horizon $R=R_g$ by
the bubble surface. As a result, in the reference frame of the distant
observer the bubble
wave function $\Psi (R,t)$ is equal to zero in the inside
region $R<R_g$ \cite{Kuch05}. Hence, only the classical trajectory $R(t)$
between the turning point $R_0$ and the gravitation radius $R_g$ can
contribute to the bubble decay rate. Energy loss by the bubble surface
becomes substantial only when $R(t)\lesssim R_0^{2/3}$ (see Appendix A
below). In our case $R_g\gg R_0^{2/3}$ and, therefore, the region of
intensive energy dissipation is not accessible by the classical trajectory.
As a result, the energy emission is negligible yielding long-lived
stationary states. In Appendix A we estimate the life-time
of a quantum bubble as a time
of energy loss by the classical bubble with the radius $R(t)$ oscillating
between $R_0$ and $R_g$; for an order of magnitude estimate we take the
oscillation period to be $R_0/c$. The answer is given by Eq. (\ref{s9})
which in dimension units reads
\begin{equation}
\label{lt}t\sim \frac{z_0^6(z_0+2)^6}{(z_0+1)^{12}}\frac{R_0}c\left( \frac{%
R_0}l\right) ^2,
\end{equation}
where $l=\hbar /mc$ is the surface width and $z_0$ is the bubble redshift at
$R=R_0$. For an axion bubble with $z_0=0.3$, $R_0=10^2R_{\odot }$ and $l<0.7$
cm Eq. (\ref{lt}) yields $t\gtrsim 10^{18}$ yrs which is much larger then
the age of the Universe.

Apart from stabilization, the quantum effects lead to broadening of emission
lines. Let us consider an emission line with the wavelength $\lambda _{0}$
produced by atoms inside the bubble. An observer outside the bubble detects
the line at the wavelength $\lambda =\lambda _{0}(1+z)$, where $z=1/\sqrt{%
1-2M/R}-1$ is the redshift of the bubble interior. At a given bubble mass $M$
the redshift depends on the radius $R$ which for the quantum bubble becomes
uncertain. Its distribution is determined by the square of the bubble wave
function $|\Psi (R)|^{2}$. Hence, the emission line is detected at the
redshift $z(R)$ with a probability proportional to $|\Psi (R)|^{2}$ and,
instead of a sharp line, the observer would detect a broaden peak centered
near $z(R)$ corresponding to a maximum of $|\Psi (R)|^{2}$. In the
quasiclassical (WKB) approximation the wave function in the interior
coordinate frame can be written as follows \cite{Auri90,Land01}
\begin{equation}
\label{w1}\Psi (R)=\frac{C}{\sqrt{p(R)}}\sin \left( \int_{R}^{R_{0}}p(x)dx+%
\frac{\pi }{4}\right)
\end{equation}
is a solution to the left of the classical turning point $R_{0}$. In the
limit of weak gravity ($uR\ll 1$) the normalization constant is $C\simeq 1.24%
\sqrt{E/R_{0}}$. In the classical forbidden region $R>R_{0}$ the wave
function decreases as
\begin{equation}
\label{w2}\Psi (R)=\frac{C}{2\sqrt{|p(R)|}}\exp \left(
-\int_{R_{0}}^{R}|p(x)|dx\right) .
\end{equation}
Near the turning point $p(R)\propto \sqrt{R_{0}-R}$. For $R>R_{0}$, $\Psi
(R) $ decays fast on the scale $\Delta R\sim 1/\alpha ^{2/3}$, hence, in the
classical forbidden region $\Psi (R)\approx 0$. For $R<R_{0}$, $\Psi (R)$
undergoes fast oscillations. Averaging over the oscillations yields in the
vicinity of the turning point
\begin{equation}
\label{w3}|\Psi (R)|^{2}\propto \frac{\Theta (R_{0}-R)}{\sqrt{R_{0}-R}},
\end{equation}
where $\Theta (x)$ is the step function. The wave function (\ref{w3}) has a
peak at $R=R_{0}$ and, therefore, the external observer would detect a peak
in the bubble radiation spectrum at the wavelength $\lambda ^{\prime
}=\lambda _{0}(1+z(R_{0}))$ as if the emission line is redshifted according
to the classical mass-radius relation (\ref{b16g}). Hence, for the quantum
bubbles the same
classical Eqs. (\ref{b16g}), (\ref{b17}) determine the redshift
of the emission features. However, the detected emission line profile $%
F(\lambda )$ would be different from those emitted by the atom. If for the
atom $F(\lambda )\propto \delta (\lambda -\lambda _{0})$ then the external
observer would detect

\begin{equation}
\label{pr}F(\lambda )\propto \frac{\Theta (\lambda -\lambda ^{\prime })}{%
\sqrt{\lambda -\lambda ^{\prime }}}.
\end{equation}
The line profile becomes asymmetric. However, the line width due to quantum
broadening is negligible and other broadening mechanisms, e.g., atom motion,
would probably wash out the quantum profile.

Finally we emphasize the dramatic difference between the classical and
quantum descriptions. In the classical picture the bubble surface localizes
after some time
at the event horizon which leads to black hole formation. However, due to
the uncertainty principle, quantum mechanics does not allow localization of
the bubble surface. Eq. (\ref{rtg}) suggests that in the reference frame of
a distant observer the bubble wave function in the WKB approximation at $%
R\rightarrow R_g$ behaves as
\begin{equation}
\label{w4}|\Psi (R)|^2\propto \frac{\Theta (R-R_g)}{R-R_g}.
\end{equation}
The WKB approximation becomes invalid in a small vicinity of $R_g$ and the
exact wave function remains finite. The normalization integral $\int |\Psi
(R)|^2dR$ of the quasiclassical wave function logarithmically diverges at $%
R\rightarrow R_g$. However, the very slow (logarithmic) divergence suggests
that if we cut off the integration at any reasonable distance to $R_g$ the
total contribution from the vicinity of $R_g$ is not very large. Hence,
in the quantum description there is quite large probability that the bubble
radius is substantially greater than $R_g$. So, in contrast to the classical
picture, the long-lived quantum bubble does not behave as a black hole.
Light escapes from the bubble interior, but it possesses gravitation
redshift determined by Eqs. (\ref{b16g}), (\ref{b17}).
This is our model of axionic quasars.

One might expect that the $1/(R-R_g)$ divergence of the WKB wave function (%
\ref{w4}) should produce features in the bubble radiation spectrum at low
frequencies. However, the redshift $z$ reduces the radiation power by a
factor of $1/(1+z)^2=(1-R_g/R)$ \cite{Misn98}. This compensates the wave
function divergence and yields no emission features at low frequencies.

\section{The nature of bright quasars}

The results obtained show that some fraction of quasars is probably bubbles
of scalar field with periodic interaction potential. They are born in active
galaxies and ejected into surrounding space. The ejection suggests that the
bubble mass must be much smaller then the mass of the parent galaxy. A
bubble with a mass $10^9$M$_{\odot }$ and gravitational redshift $z=1$ has a
radius $R=10^{-4}$pc and, if located at a distance $1$ Mpc, would have an
angular size of $0.00004^{\prime \prime }$ which is well below the
resolution limit of current telescopes in the optical band ($0.1^{\prime
\prime }$). Therefore, the dark matter bubbles can explain only the faint
quasars with unresolved structure. Allowed values of the mass and decay
constant $f$ of axion fit well into this picture suggesting the axion as
primary candidate for composition of the dark matter bubbles.

The question whether or not all dark matter in the Universe consists of
axions requires further detailed analysis. However, existence of bright
quasars with resolved ``host'' galaxies suggests that some part of dark
matter is composed of other particles. Fuzz around bright quasars 3C 48 ($%
z=0.37$) \cite{Cana00}, 3C 273 ($z=0.158$) \cite{Hipp96}, 3C 279 ($z=0.536$)
\cite{Cheu02} was found with the redshift equal to the quasar redshift. The
angular size of the fuzz is $0.5^{\prime \prime }$ (3C 279), $4^{\prime
\prime }$ (3C 48) and $10^{\prime \prime }$ (3C 273). Due to large angular
size and brightness such quasars can not be axionic bubbles. However, based
on observations, Arp has suggested that 3C 273 and 3C 279 are members of the
Virgo cluster ($21$ Mpc away), while 3C 48 belongs to the Local Group \cite
{Arp98}. There is considerable evidence from the morphology \cite{Arp90a}
and from X-ray emission \cite{Arp94} that this may be the case and, hence,
those objects possess noncosmological redshift.

Dark matter bubbles or droplets with radii $10-500$ pc can explain the
observed angular size of the bright quasars provided that the redshift is
gravitational. However, if we estimate the required mass from Eq. (\ref{b17}%
) we obtain $10^{14}-10^{15}$M$_{\odot }$ which is much greater than a
typical mass of a large galaxy $10^{12}$M$_{\odot }$. Such big masses
definitely would not fit into any realistic picture of galaxy clusters and
contradict to the scenario of quasar ejection from galaxies.

Here we discuss a solution of the Einstein-Klein-Gordon equations describing
an object with galactic size, large gravitational redshift, but possessing
negligibly small mass compared to the mass of a galaxy. The solution is
similar to a spherical capacitor known in electrostatic. If one plate of the
``capacitor'' has the mass $+M$, while the other one possesses the mass $-M$
the total mass and gravitational field outside the capacitor is equal to
zero. However, the capacitor interior has a nonzero gravitational potential
and, hence, nonzero gravitational redshift. Such a capacitor can be realized
for scalar fields with negative (attractive) interaction potential. The
field kinetic energy provides positive contribution to the energy density,
while the contribution from interaction is negative. Space regions where the
kinetic energy dominates play a role of the plates with positive mass, while
interaction dominating regions are analogous to the negative plates. One can
anticipate the effect from the structure of the Einstein equation (\ref{b6})
for the metric component $N$ that determines the space redshift $z=1/N-1$.
If $V<0$ then the combination $\varphi ^{\prime 2}-2g^2V$ everywhere gives
the positive contribution to $N$ producing large total redshift. However,
according to Eq. (\ref{e3}), the mass density depends on $\varphi ^{\prime
2}+2g^2V$ and if $V<0$ the total mass can be very small.

To demonstrate the effect we consider a complex scalar field $\psi $ with
the simplest attractive potential
\begin{equation}
\label{n1}V(\psi )=-m^2|\psi |^2/2.
\end{equation}
The free field describes tachyons that always propagate with the speed
greater then the speed of light $c$, possess momentum $p\geq mc$ and the
energy $E=\sqrt{p^2c^2-m^2c^4}$. There is an evidence that neutrino could be
a tachyon \cite{Chod85,Moha98}, which makes the hypothesis that dark matter
is partially composed of tachyons not so exotic. Tachyons as a candidate for
dark matter and dark energy are currently discussed in the literature (see,
e.g., \cite{Shiu02,Bagl03,Davi04} and references therein). They can explain
the observations at both large and galactic scales \cite{Padm02,Caus04}.

The equilibrium field configurations are those in which the metric is time
independent. The scalar field $\psi $ itself can oscillate with frequency $%
\omega $, $\psi (t,r)=\psi (r)\exp (-i\omega t)$, however due to U(1)
symmetry of the Lagrangian the space-time geometry is static. In
dimensionless units, defined by Eq. (\ref{a2}), the stationary Klein-Gordon
and Einstein equations for $\psi (r)$ are \cite{Seid90}
\begin{equation}
\label{bn5}\frac{\psi ^{\prime }}{g^2}\left( \frac{g^2+1}r+rg^2\psi
^2\right) +\frac{\psi ^{\prime \prime }}{g^2}+\frac{\omega ^2\psi }{m^2N^2}%
+\psi =0,
\end{equation}
\begin{equation}
\label{bn6}N^{\prime }=\frac N2\left[ \frac{g^2-1}r+r\left( \psi ^{\prime
2}+g^2\psi ^2+\frac{\omega ^2g^2\psi ^2}{m^2N^2}\right) \right] ,
\end{equation}
\begin{equation}
\label{bn7}g^{\prime }=\frac g2\left[ \frac{1-g^2}r+r\left( \psi ^{\prime
2}-g^2\psi ^2+\frac{\omega ^2g^2\psi ^2}{m^2N^2}\right) \right] .
\end{equation}
For simplicity we assume $\omega =0$. In the limit of zero gravity the
Klein-Gordon equation
\begin{equation}
\label{n2}\Delta \psi +\psi =0
\end{equation}
has a spherically symmetric solution
\begin{equation}
\label{n3}\psi =\psi _0\frac{\sin r}r
\end{equation}
that describes a droplet with specially oscillating energy density $E(r)$
which, at $r\gg 1$, is given by $E(r)=\psi ^{\prime 2}/2-\psi ^2/2=\psi
_0^2\cos (2r)/2r^2$. The solution is analogous to a set of concentric
spherical plates charged alternatively with positive and negative mass.

General relativistic treatment modifies the formula. First we discuss the
solution of Eqs. (\ref{bn5})-(\ref{bn7}) in the limit $|\psi |\ll 1$. In
such limit one can take $g\approx 1$, then Eq. (\ref{bn5}) leads to
\begin{equation}
\label{n4}\psi ^{\prime }\left( \frac 2r+r\psi ^2\right) +\psi ^{\prime
\prime }+\psi =0.
\end{equation}
Further we take the nonlinear term $r\psi ^2$ in the zero order
approximation (\ref{n3}). This term is important at $r\gg 1$ and yields a
decay of the scalar field faster then $1/r$. Also we average the term $r\psi
^2$ over the period of space oscillation, that is assume $r\psi ^2\approx
\psi _0^2/2r$, and finally obtain
\begin{equation}
\label{n5}(2+\psi _0^2/2)\psi ^{\prime }/r+\psi ^{\prime \prime }+\psi =0.
\end{equation}
Solution of this equation is expressed in terms of the Bessel function
\begin{equation}
\label{n6}\psi =\frac{\sqrt{\pi }\psi _0J_{1/2+\psi _0^2/4}(r)}{\sqrt{2}%
r^{1/2+\psi _0^2/4}}
\end{equation}
and has the following asymptotics: $\psi \approx \psi _0$ at $r\ll 1$, and $%
\psi \approx \psi _0\sin r/r^{1+\psi _0^2/4}$ at $r\gg 1$. The total
mass-energy of the droplet is (in units $m_{\text{pl}}^2/m$)
\begin{equation}
\label{n7}E=\int_0^\infty r^2\left( \frac{\psi ^{\prime 2}}{g^2}-\psi
^2\right) dr \lesssim \psi _0^4,
\end{equation}
which for $\psi _0\ll 1$ is negligible. However, for the gravitational
redshift at the droplet center $z_c$ we obtain
\begin{equation}
\label{n8}z_c\approx \exp \left[ \frac 12\int_0^\infty r(\psi ^{\prime
2}+\psi ^2)dr\right] -1=e-1=1.718,
\end{equation}
the value is independent of $\psi _0$ (in the limit $\psi _0\ll 1$). If we
move away from the droplet center the redshift of the space decreases from $%
z_c$ to zero with the asymptotic behavior $z(r)=1/r^{\psi _0^2/2}$ ($r\gg 1$%
).

\begin{figure}[tbp]
\bigskip
\centerline{\epsfxsize=0.45\textwidth\epsfysize=0.33\textwidth
\epsfbox{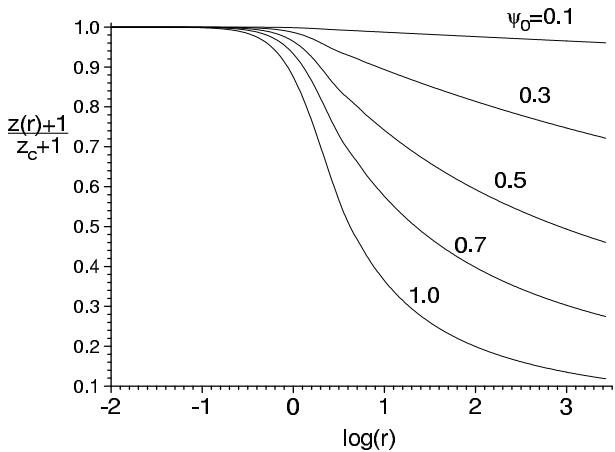}}
\par
\caption{Redshift $z(r)$ of space as a function of distance $r$ to the
droplet center. Note that the redshift is normalized to its center value $z_c $ and the scale along the horizontal axis is logarithmic. The unit of
length is $\hbar/mc$. }
\label{zcf}
\end{figure}

Numerical solution of Eqs. (\ref{bn5})-(\ref{bn7}) shows that depending on
the scalar field at the droplet center $\psi _0$ the gravitational redshift
can have any large values. In Fig. \ref{zcf} we plot the spacial
distribution of the redshift $z(r)$ for different values of $\psi _0$. Near
the droplet center the redshift changes on the scale of a few $\hbar /mc$,
while far from the center the spacial variation becomes very weak.

Our model can provide an estimate of the characteristic scales. If, e.g., we
take $m=0.64\times 10^{-23}$ eV then the unit of length is $\hbar /mc=1$ pc.
A bright baryonic nucleus located a distance $r=10$ kpc from the center of
the tachyonic droplet with $\psi _0=0.2$ (the droplet energy $|E| \lesssim
\psi_0^4m_{\text{pl}}^2/m=10^{10}M_{\odot }$) would possess a redshift $%
z=1/r^{\psi _0^2/2}=0.83$. Fuzz surrounding the baryonic nucleus with a
diameter $\Delta r=200$ pc would have a dispersion of the gravitational
redshift $\Delta z\approx \psi _0^2\Delta r/2r^{1+\psi _0^2/2}=0.00033=100$
km/s, the redshift increases towards the droplet center. Such systematic
change in the redshift across the fuzz can be directly measured and serve as
a test of our theory. However, the position of the bright nucleus might also
coincide with the droplet center, such a picture can be realized provided we
choose for $m$ a smaller value.

Our finding shows that substantial gravitational redshift of a large space
volume does not necessarily require presence of a big mass. Even a volume of
galactic scale can possess any value of the gravitational redshift produced
by dark matter with the total mass negligibly small compared to galactic
mass. Ordinary baryonic matter placed in such a volume behaves as being
possessed of an intrinsic redshift. The effect can explain the nature of
intrinsically bright quasars as nuclei of forming nearby small galaxies
embedded in a clot of scalar field with negative interaction. It also can
account for recent observation of the field surrounding the Seyfert galaxy
NGC 7603 where four galaxies with substantially different redshifts are
apparently connected by a narrow filament \cite{Lope04}.

We want to emphasize that the interaction potential (\ref{n1}) provides a
simple demonstration of the effect, rather then a precise quantitative
description of the phenomena. Realistic potential must be derived based on
detailed study of bright quasars including their environment and, probably,
should give rise to a faster asymptotic decay of the scalar field with the
distance from the droplet center.

\section{Discussion}

\subsection{Axions}

In this paper we argue that the problems of quasars and dark matter are
mutually related. We found that bubbles of scalar field with periodic
interaction potential can explain the nature of intrinsically faint
point-like quasars associated with nearby galaxies. Typical absolute
magnitude $M_v$ of such quasars lies in the range $-8$ to $-13$ (optical
luminosity $L=10^5-10^7L_{\odot }$) \cite{Burb03,Arp04b}, the redshift is
mostly gravitational and found to be quantized. The bubbles are born in
nuclei of active galaxies and ejected into surrounding space. They cluster
at a distance upto 100$-$300 kpc from the parent galaxy. About 15 such
objects have been discovered close to M82, the nearest active galaxy to the
Milky Way \cite{Burb03}, and about 10 in the vicinity of NGC 3628 \cite
{Arp02}.

The observed five peaks in the quasar redshift distribution match well the
theoretical result with only one free parameter, which is a strong argument
in favor of our theory. The hypothetical axions, one of the leading dark
matter candidate, fit well into the quasar picture and can account for the
bubble composition. The axion mass range constrained by astrophysical and
cosmological arguments yields the necessary bubble life-time and the size
which agrees with the observed quasar brightness.

\begin{figure}[tbp]
\bigskip
\centerline{\epsfxsize=0.45\textwidth\epsfysize=0.25\textwidth
\epsfbox{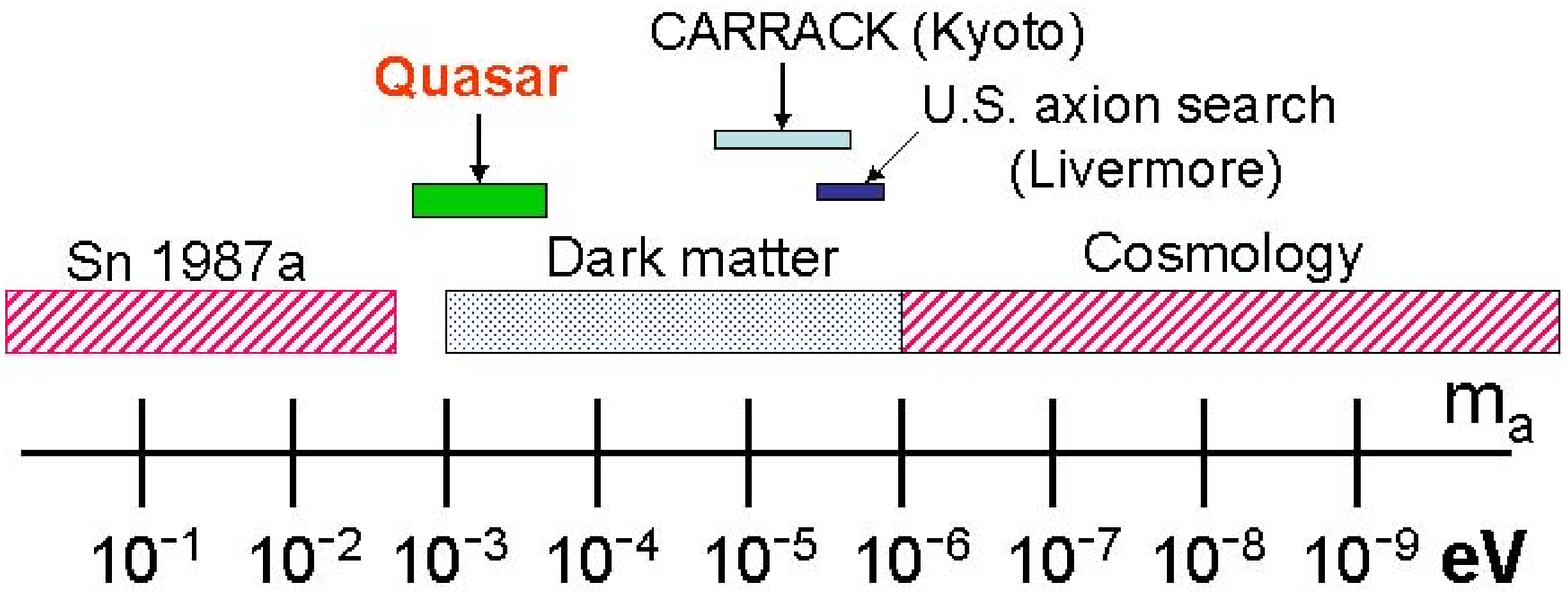}}
\caption{Astrophysical and cosmological exclusion regions (hatched) for the
axion mass $m_a$. The dotted ``inclusion region" indicate where axions could
plausibly be the cosmic dark matter. Axion mass determined from quasar
observations fits in the open window. Also shown is the projected
sensitivity range of the search laboratory experiments for galactic dark
matter axions. }
\label{mass}
\end{figure}

Properties of the intrinsically faint QSOs, combined with equations for the
bubble mass $M=0.00752\alpha ^2m_{\text{pl}}^2/m=2.94m\text{(eV)}\times
10^{11}M_{\odot }$, radius $R=0.0329\alpha ^2\hbar /mc=2.73m\text{(eV)}%
\times 10^6R_{\odot }$ and the relation (\ref{g1}) $m=0.62$ eV$\times 10^7$%
GeV/$f$ \cite{Brad03}, allow us to determine the axion mass $m$. The quasar
luminosity suggests that the bubble radius is larger then $10^3R_{\odot }$
which yields $m>4\times 10^{-4}$eV and $M>10^8M_{\odot }$. From the other
hand, the quasar ejection from active galaxies implies that the bubble mass $%
M$ must be much smaller then the galactic mass. It is reasonable to
constrain $M<10^9M_{\odot }$ which leads to $m<3\times 10^{-3}$eV and $%
R<10^4R_{\odot }$. We conclude, the axion mass is $m=0.4-3$ meV. This value
fits in the open window for the axion mass constrained by astrophysical and
cosmological arguments, as displayed in Fig. \ref{mass}, which unambiguously
points towards the axionic nature of dark matter composing the intrinsically
faint point-like quasars. One can see that current cavity axion search
experiments in Livermore \cite{Aszt02,Aszt04} and Kyoto University \cite
{Brad03} are looking for the axion in an unlikely mass range which deviates
by two orders of magnitude from our result (see Fig. \ref{mass}). Probably
now, when the axion mass is established from quasar observations, the axion
has a better chance to be discovered. One should mention that radio
telescopes are suited to search for axions of higher mass \cite{Blou01} and
allow searches in the range obtained in our paper.

Understanding the mechanisms of quasar formation requires further detailed
study. Bose-Einstein condensation of axions in galactic centers can be a
possible natural mechanism of bubble creation. Dark matter axions, if they
exist, form halos around galaxies. The halo axions are in a quantum
degenerate non-equilibrium regime with the de Broglie wavelength $\lambda
_d\sim 1$ m which is much larger then the interparticle spacing \cite{Brad03}%
. The ground state of the system would be a collapsed axion condensate at
the galactic center. However, the kinetics of condensate nucleation is
governed by two factors: inter-particle collisions and occupation number of
the condensate particles. Due to extremely weak interactions between axions
the collision rate is very small, so the galactic halo remains metastable
for a long time.

Recent experiments on Bose condensate formation in magnetic traps \cite
{Kohl02} show that in a weak cooling regime the condensate nucleation occurs
in two stages: slow linear growth which then, at a critical point, triggers
fast exponentially growing instability, caused by the effect of bosonic
stimulation. Similar situation might take place in nuclei of active
galaxies. At the first stage, the axions are slowly (during million years)
accumulated at the central part of the galaxy. When the mass of such a ball
becomes critical (particle occupation number is big enough), it triggers
exponential instability due to bosonic stimulation. A dense condensate
cloud, coherent on astronomically large scale, starts to form rapidly. The
cloud then collapses under its own gravity. Such a mechanism is similar to
type Ia supernovae, when a star explodes after accumulation a critical mass,
and suggests formation of objects with approximately equal masses. Why does
the cloud collapse lead to formation of bubbles? The answer comes from a
three-dimensional numerical simulation of the evolution of inhomogeneities
in the axion field. Such a simulation has indeed demonstrated formation of
bubble-like structures (see Fig. 5a in Ref. \cite{Kolb94}). Gravitational
cooling is probably an important processes involved in quasar nucleation
\cite{Seid94}.

Finally we want to mention a new (preliminary) experimental result which
became available after our theory was first presented in DARK 2004 \cite
{Svid04}. The PVLAS experiment on axion laser production and detection
apparently observes a signal which might be produced by axion-like particles
\cite{PVLAS}. If the observed signal is indeed caused by axions then the
particle mass measured in the PVLAS experiment is $m=1.0\pm 0.1$ meV \cite
{PVLAS}. This value is right in the middle of the axion mass interval $0.4-3$
meV we predicted based on quasar observations! One should mention that
another parameter, the axion-photon coupling, measured by PVLAS is
irrelevant to our theory. Our estimate of the axion mass is based only on
the relation (\ref{g1}) between the global symmetry-breaking scale $f$ and $m
$. From this perspective one can treat our and PVLAS results as mutually
complementary. Only combination of the two independent results confirm,
based on observations, the relation (\ref{g1}) which is the key ingredient of
any axion model. This provides an evidence that the new particle, detected by
PVLAS and responsible for the quasar redshift quantization, is indeed axion and
not another yet unknown pseudoscalar particle.

\subsection{Tachyons}

Existence of the both intrinsically faint point-like and bright resolved
quasars associated with galaxies suggests that dark matter in the Universe
is probably composed of several species. We found that tachyonic clots can
produce substantial gravitational redshift in a galactic scale even when the
total mass of the object is negligible. Bose condensate of tachyons can
explain the nature of intrinsically bright quasars as forming galactic
nuclei emerged into droplets of scalar field with negative interaction. The
droplets possess large gravitational redshift on kpc scales. About 42 such
objects have been discovered in the vicinity of the Seyfert 1 galaxy NGC
6212 which is approximately 120 Mpc away \cite{Burb03a}. They form a cloud
of QSOs with a characteristic size $2$ Mpc and absolute QSO magnitudes from $%
-20$ to $-15$.

Being large and luminous, the tachyonic objects dominate in samples of
quasars associated with relatively distant, $\gtrsim 50$ Mpc, active
galaxies. At such distances the axionic bubbles are so faint that they are
rarely detectable as individual sources. According to our theory, the
redshift quantization is the property of axionic bubbles and must not be
present for bright tachyonic quasars. In agreement with the theory, no
periodicity in the intrinsic redshift distribution has been found in a
sample of tachyon dominating quasars associated with the distant ($0.01<z_{%
\text{gal}}<0.3$) galaxies \cite{Hawk02}.

If within a distance $d\lesssim 10$ Mpc there are, at least, several bright
tachyonic quasars, then inside the region $d\lesssim 100$ Mpc thousands of
them must be detected. Probably such objects dominate in the recent First
Bright Quasar Survey which radio-selected sources brighter than 18 optical
magnitude \cite{Whit00,Beck01}; axionic bubbles are usually too faint and
not selected in such a survey. Luminous tachyonic quasars can be also seen
at cosmological distances where the cosmological redshift contribution
becomes substantial. The gravitational redshift of tachyonic quasars can
also be very small which provides a continuous connection between quasars
and active galactic nuclei, the classical quasar explanation. In general
case, however, the quasar redshift possesses substantial gravitational
component and, hence, is not a measure of the object distance.

Proximity of some quasars and galaxies with approximately equal redshift
indicates only that for some objects the gravitational redshift is small
compared to cosmological. Meanwhile, the statistical evidence, mentioned in
the Introduction, implies that the gravitational redshift dominates for the
majority of quasars in the selected sample. It is also worth to note that if a
dwarf galaxy lies at a distance of a few kpc from the tachyonic quasar, both
such objects are emerged into the same tachyonic clot and, as a consequence,
must have close redshifts. This could be a reason why some quasars and
associated galaxies possess equal redshifts and account for the nature of
binary quasars.

\begin{figure}[tbp]
\bigskip
\centerline{\epsfxsize=0.33\textwidth\epsfysize=0.31\textwidth
\epsfbox{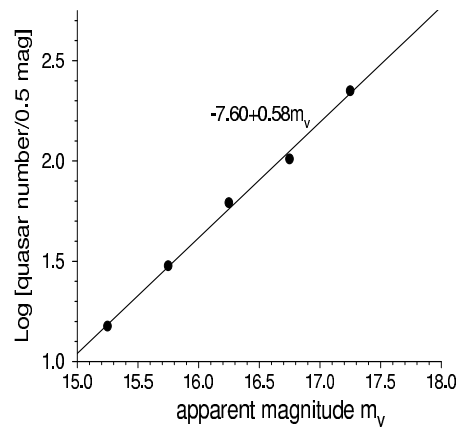}}
\caption{Logarithm of the number of quasars within the $\Delta m_v=0.5$
magnitude bins in the First Bright Quasar Survey of the north Galactic cap
as a function of the apparent magnitude $m_v$ (dots) \protect\cite{Whit00}.
The solid line is the best linear fit.}
\label{first}
\end{figure}

It is known that quasars violate the Hubble redshift-apparent magnitude
relation which is an argument against the cosmological redshift. The
argument would be very strong if the quasar apparent magnitude $m_v$ proved
to be a good indicator of their distance. If objects with equal luminosity
possess a uniform space distribution then their number $N$ detected per unit
$m_v$ obeys the law $\log (dN/dm_v)=const+0.6m_v$ \cite{Sone77}. In Fig. \ref
{first} we plot $\log (dN/dm_v)$ as a function of the optical magnitude $m_v$
for quasars discovered by the First Bright Quasar Survey (dots) \cite{Whit00}%
. The solid line is the best linear fit which yields the slope of $0.58$
very close to $0.6$. Such an agreement suggests that for a large sample
mostly the distance determines $m_v$ of the bright quasars and, hence, the
scattered Hubble $z-m_v$ diagram implies noncosmological redshift for the
majority of quasars in the Survey.

Arp has noticed an empirical sequence of quasar development in which
initially point-like objects at relatively faint apparent magnitudes
transform into lower-redshift compact objects with ``fuzz'' around their
perimeters, and then into small, high surface brightness galaxies \cite
{Arp97,Arp98a,Arp98}. In the light of our theory this means that faint
axionic bubbles, which are born in and ejected from active galaxies, later
serve as nucleation centers for tachyon Bose-Einstein condensation.
Condensate formation leads to appearance of a large in size, bright quasar
which then gradually transforms into a small companion galaxy. Tachyonic
clots can account for the systematic excess of the redshift of small
companion galaxies compared to the redshift of the dominant galaxy \cite
{Arp98}. Continuity suggests that the dark halos of galaxies, known from
their rotation curves, must be remnants of the tachyonic matter. Possibility
of such scenario is a subject of future study.

The process of axionic quasar nucleation could produce substantial
disturbances in galaxies. Such catastrophic events can contribute to the
observed gamma-ray bursts. However, existence of nearby quasars suggests
that such events also occur in our epoch, not only in the early Universe. If
we divide the total number of galaxies in the Universe, $8\times 10^{10}$,
by the frequency of the gamma-ray bursts ($\sim 1$ per day) we obtain that
in average the catastrophic event occurs with the interval of $100$ million
years per galaxy. This value is of the order of the time span between
geological periods on Earth which might indicate on their Galactic origin.
More often catastrophic events occurring in the Local galactic group might
be responsible for the change in the geological ages within a period. Study
of the geological periods on other planets, e.g., on Mars, can verify this
hypothesis.

I wish to thank A. Belyanin and A. Sokolov for helpful discussions.

\appendix

\begin{widetext}
\onecolumngrid

\section{Energy emission from a shrinking bubble}

Here we calculate energy loss by a shrinking spherically symmetric bubble
caused by emission of scalar particles. For an order of magnitude estimate
one can omit the effect of gravity. Then the evolution of the scalar field $%
\varphi (t,r)$ is described by sine-Gordon equation
\begin{equation}
\label{s1}\ddot \varphi -\varphi ^{\prime \prime }+\sin \varphi =2\varphi
^{\prime }/r,
\end{equation}
where $r$ is the radial coordinate. Without right-hand side, Eq. (\ref{s1})
has an exact, so-called kink, solution
\begin{equation}
\label{s2}\varphi _0=4\arctan \left\{ \exp \left[ \pm \frac{(r-vt-R_0)}{\sqrt{1-v^2}}\right] \right\} ,
\end{equation}
where $R_0\gg 1$ is the initial bubble radius. The solution describes a kink
(space region where $\varphi $ changes from $2\pi $ to $0$) propagating with
constant velocity $v$; the kink's size is $l\sim \sqrt{1-v^2}$.

If $l\ll R(t)$, where $R(t)$ is the bubble radius, r.h.s. of (\ref{s1}) may
be treated as a small perturbation. Eq. (\ref{s1}) possesses approximate
solution in the form of the kink (\ref{s2}) with parameters slowly changing
in time under the action of the perturbation. In particular, the kink
shrinks due to its surface tension so that the bubble radius and the
velocity evolve as \cite{Malo87a}
\begin{equation}
R(t)=R_0cn(\sqrt{2}t/R_0,1/\sqrt{2})\text{, \ }v(t)=\sqrt{1-R^4(t)/R_0^4}%
\text{,}
\end{equation}
where $cn$ stands for the elliptic cosine with the modulus $1/\sqrt{2}$.
Such a process is accompanied by emission of scalar particles which yields
the energy loss. We estimate the energy loss following the original work of
Malomed \cite{Malo87a,Malo87b}. In terms of the inverse scattering
technique, the spectral density of the emitted energy $E_e(t,q)$ is

\begin{equation}
\label{s3}\frac{dE_e}{dq}=\frac 4\pi |B(t,q)|^2,
\end{equation}
where $q$ is the radiation wavenumber and the perturbation-induced evolution
equation for the complex amplitude $B(t,q)$ is given by \cite
{Malo87a,Malo87b}
\begin{equation}
\label{s4}\frac{dB}{dt}=-\frac i{2(\lambda ^2+\gamma ^2)}\int_{-\infty
}^\infty dr\left( \lambda ^2-\gamma ^2-2i\lambda \gamma \tanh \left[ \frac{r-vt}{\sqrt{1-v^2}}\right] \right) \exp \left( i\sqrt{1+q^2}t-iqr\right)
\partial _r\varphi _0,
\end{equation}
where $\lambda =\sqrt{1+q^2}-q$ and $\gamma =(1+v)/2\sqrt{1-v^2}$.
Calculating the integral in (\ref{s4}) yields
\begin{equation}
\label{s5}\frac{dB}{dt}=\frac{i\pi \left[ \lambda ^2(1-v)(1-\sqrt{1+v})-v/2\right] }{(1+v)/4+\lambda ^2(1-v)}\frac{\exp \left( i\sqrt{1+q^2}t-iqvt\right) }{\cosh \left[ \pi q\sqrt{1-v^2}/2\right] }
\end{equation}
If $v$ slowly varies with time one can take $v\approx const$ in Eq. (\ref{s5}%
), then after integration we obtain
\begin{equation}
B(t,q)=\frac{\pi \left[ \lambda ^2(1-v)(1-\sqrt{1+v})-v/2\right] }{\left[
(1+v)/4+\lambda ^2(1-v)\right] (\sqrt{1+q^2}-qv)}\frac{\exp \left( i\sqrt{1+q^2}t-iqvt\right) -1}{\cosh \left[ \pi q\sqrt{1-v^2}/2\right] }.
\end{equation}
Therefore
\begin{equation}
\label{s6}\frac{dE_e}{dq}=\frac{16\pi \left[ \lambda ^2(1-v)(1-\sqrt{1+v})-v/2\right] ^2}{\left[ (1+v)/4+\lambda ^2(1-v)\right] ^2(\sqrt{1+q^2}-qv)^2}%
\frac{\sin ^2\left[ \left( \sqrt{1+q^2}-qv\right) t/2\right] }{\cosh
^2\left[ \pi q\sqrt{1-v^2}/2\right] }.
\end{equation}
Integration of (\ref{s6}) over $dq$ gives the emitted energy as a function
of time $E_e(t)=\int_{-\infty }^\infty dq(dE_e/dq)$. In Eq. (\ref{s6}) sine
is a fast oscillating function, so we substitute $\sin ^2(x)\rightarrow 1/2$%
. The radiation power increases when the kink's velocity $v$ approaches the
speed of light $c=1$. Assuming $1-v\ll 1$, integration of Eq. (\ref{s6})
yields
\begin{equation}
\label{s7}E_e(t)\approx \frac{2.51}{(1-v(t))^{3/2}}\approx 7.10\left( \frac{R_0}{R(t)}\right) ^6
\end{equation}
The emitted energy becomes comparable with the initial bubble energy $%
E_0=8R_0^2$ when the bubble radius reaches the value $R_{*}\approx R_0^{2/3}$%
. This value agrees with those obtained in \cite{Widr89}.

If the bubble shrinks from $R_0$ to the gravitation radius $R_g\gg R_{*}$
the radiated energy is
\begin{equation}
\label{s8}E_g\sim 7.10\left( \frac{R_0}{R_g}\right) ^6\approx \frac{(z_0+1)^{12}}{z_0^6(z_0+2)^6}\frac{E_0}{R_0^2}\text{,}
\end{equation}
where $z_0=1/\sqrt{1-R_g/R_0}-1$ is the initial bubble redshift. To emitt
all its energy the bubble must oscillate between $R_g$ and $R_0$ about $%
E_0/E_g$ cycles. As a result, the bubble life-time is
\begin{equation}
\label{s9}t\sim R_0\frac{E_0}{E_g}\approx \frac{z_0^6(z_0+2)^6R_0^3}{(z_0+1)^{12}},
\end{equation}
which for $z_0=0.3$ yields $t\sim 0.005R_0^3$.

\end{widetext}\twocolumngrid


\begin{thebibliography}{999}
\bibitem{Schm63}  M. Schmidt, Nature \textbf{197}, 1040 (1963).

\bibitem{Sand73}  A.R. Sandage, ApJ \textbf{180}, 687 (1973).

\bibitem{Arp87}  H. Arp, \textit{\ Quasars, redshifts and controversies. }
Interstellar Media, Berkeley, 1987.

\bibitem{Narl89}  J.V. Narlikar, Space Sci. Rev. \textbf{50}, 523 (1989).

\bibitem{Burb92}  G. Burbidge and A. Hewitt,
\textit{In Variability of blazars}, Eds E. Valtaoja and M. Valtonen,
Cambridge University Press, Cambridge, p. 4 (1992).

\bibitem{Arp98}  H. Arp,
\textit{Seeing red: redshifts, cosmology and
academic science. }Apeiron, Montreal, 1998.

\bibitem{Kemb99}  A. K. Kembhavi and J. V. Narlikar 1999,
\textit{Quasars and
Active Galactic Nuclei}. Cambridge University press, Cambridge, 1999, Sec.
15.

\bibitem{Burb01p}  G. Burbidge, PASP, \textbf{113}, 899 (2001).

\bibitem{Bahc94}  J. Bahcall, S. Kirkakos, and D. Schneider, ApJ \textbf{435}%
, L11 (1994).

\bibitem{Hutc95}  J.B. Hutchings and S. Morris, AJ \textbf{109}, 1541 (1995).

\bibitem{Stoc78}  A. Stockton, Nature \textbf{274}, 342 (1978); ApJ
\textbf{223}, 747 (1978).

\bibitem{Heck84}  T.M. Heckman, G.D. Bothun, B. Balick and E.P. Smith, AJ
\textbf{89}, 958 (1984).

\bibitem{Yee87}  H.K.C. Yee, AJ \textbf{94}, 1461 (1987).

\bibitem{Burb90}  G.R. Burbidge, A. Hewitt, J.V. Narlikar and P. Das Gupta,
ApJS \textbf{74}, 675 (1990).

\bibitem{Burb96}  G. Burbidge, A\&A, \textbf{309}, 9 (1996).

\bibitem{Beni97}  N. Benitez and E. Martinez-Gonzalez, ApJ \textbf{477}, 27
(1997).

\bibitem{Burb03}  E. M. Burbidge , G. Burbidge, H. Arp and S. Zibetti, ApJ
\textbf{591}, 690 (2003).

\bibitem{Arp04a}  H. Arp, E.M. Burbidge and G. Burbidge, A\&A \textbf{414},
L37 (2004).

\bibitem{Arp04b}  H. Arp, C.M. Guti\'errez and M. L\'opez-Corredoira, A\&A
\textbf{418}, 877 (2004).

\bibitem{Gali04}  P. Galianni, E.M. Burbidge, H. Arp, V. Junkkarinen, G.
Burbidge and S. Zibetti, prepint astro-ph/0409215.

\bibitem{Burb99}  E.M. Burbidge, ApJ \textbf{511}, L9 (1999).

\bibitem{Arp02}  H. Arp, E.M. Burbidge, Y. Chu, E. Flesch, F. Patat and G.
Rupprecht, A\&A \textbf{391}, 833 (2002).

\bibitem{Ferr97}  I. Ferreras, N. Benitez and E. Martinez-Gonzalez, AJ
\textbf{114}, 1728 (1997).

\bibitem{Karl90}  K.G. Karlsson, A\&A \textbf{239}, 50 (1990).

\bibitem{Arp90}  H. Arp, H.G. Bi, Y. Chu and X. Zhu, A\&A \textbf{239}, 33
(1990).

\bibitem{Burb01}  G. Burbidge and W.M. Napier, AJ \textbf{121}, 21 (2001).

\bibitem{Hawk02}  E. Hawkins, S.J. Maddox and M.R. Merrifield, MNRAS
\textbf{336}, L13 (2002).

\bibitem{Svid04}  A.A. Svidzinsky, {\it ``Intrinsically faint quasars:
evidence for meV axion dark matter in the Universe'' } in Proceedings of the
International Conference DARK 2004, College Station, 3-9 October, 2004, Eds.
H.V. Klapdor-Kleingrothaus and R. Arnowitt, Springer (2005);
astro-ph/0411548.

\bibitem{Sado99}  B. Sadoulet, Rev. Mod. Phys. \textbf{71}, S197 (1999).

\bibitem{Brad03}  R. Bradley, J. Clarke, D. Kinion, L. J. Rosenberg, K. van
Bibber, S. Matsuki, M. M\"uck and P. Sikivie, Rev. Mod. Phys. \textbf{75},
777 (2003).

\bibitem{Moor99}  B. Moor et al., MNRAS \textbf{310}, 1147 (1999).

\bibitem{Arbe03}  A. Arbey, J. Lesgourgues and P. Salati, Phys. Rev. D
\textbf{68}, 023511 (2003).

\bibitem{Mato01}  T. Matos and L.A. Ure\~na-L\'opez, Phys. Rev. D \textbf{63}%
, 063506 (2001).

\bibitem{Lope02}  L.A. Ure\~na-L\'opez and A. R. Liddle, Phys. Rev. D
\textbf{66}, 083005 (2002).

\bibitem{Fuch04}  B. Fuchs and E.W. Mielke, MNRAS \textbf{350}, 707 (2004).

\bibitem{Mato03}  T. Matos and G. Torres, RMxAA \textbf{39}, 113 (2003).

\bibitem{Frie95}  J.A. Frieman, C.T. Hill, A. Stebbins and I. Waga, Phys.
Rev. Lett. \textbf{75}, 2077 (1995).

\bibitem{Vian99}  P.T.P. Viana and A.R. Liddle, Astrophysics and Space
Science \textbf{261}, 291 (1999).

\bibitem{Hu00}  W. Hu, R. Barkana and A. Gruzinov, Phys. Rev. Lett.
\textbf{85}, 1158 (2000).

\bibitem{Silv01}  M.P. Silverman and R.L. Mallett, Class. Quantum Grav.
\textbf{18}, L103 (2001).

\bibitem{Hill02}  C.T. Hill and A.K. Leibovich, Phys. Rev. D \textbf{66},
075010 (2002).

\bibitem{Hill88}  C.T. Hill and G.G. Ross, Nuclear Phys. B\textbf{311}, 253
(1988).

\bibitem{Pecc77}  R.D. Peccei and H. Quinn, Phys. Rev. Lett. \textbf{38},
1440 (1977).

\bibitem{Raff02}  G. Raffelt, Space Science Rev. \textbf{100}, 153 (2002).

\bibitem{Kim87}  J.E. Kim, Phys. Rep. \textbf{150}, 1 (1987).

\bibitem{Seid90}  E. Seidel and W. M. Suen, Phys. Rev. D \textbf{42}, 384
(1990).

\bibitem{Rose66}  G. Rosen, J. Math. Phys. \textbf{7}, 2066 (1966);
\textbf{7}, 2071 (1966).

\bibitem{Koda78}  T. Kodama, K.C. Chung and A.F. da F. Teixeira, Nuovo
Cimento \textbf{46}, 206 (1978).

\bibitem{Karl71}  K.G. Karlsson, A\&A \textbf{13}, 333 (1971); A\&A
\textbf{58}, 237 (1977).

\bibitem{Barn76}  J.M. Barnothy and M.F. Barnothy, PASP \textbf{88}, 837
(1976).

\bibitem{Napi03}  W.M. Napier and G. Burbidge, MNRAS \textbf{342}, 601
(2003).

\bibitem{Glen00}  N.K. Glendenning, ``Compact Stars: Nuclear Physics,
Particle Physics, and General Relativity'', Springer Verlag; New York, 2nd
edition, (2000).

\bibitem{Thua79}  T. X. Thuan, J. B. Oke and J. Bergeron, ApJ \textbf{230},
340 (1979).

\bibitem{Wang03}  T.G. Wang and X.G. Zhang, MNRAS \textbf{340}, 793 (2003).

\bibitem{Auri84}  A. Aurilia, G. Denardo, F. Legovini and E. Spallucci,
Phys. Lett. \textbf{147B}, 258 (1984).

\bibitem{Blau87}  S. K. Blau, E.I. Guendelman and A.H. Guth, Phys. Rev. D
\textbf{35}, 1747 (1987).

\bibitem{Auri87}  A. Aurilia, R.S. Kissack, R. Mann and E. Spallucci, Phys.
Rev. D \textbf{35}, 2961 (1987).

\bibitem{Bere87}  V.A. Berezin, V.A. Kuzmin and I.I. Tkachev, Phys. Rev. D
\textbf{36}, 2919 (1987).

\bibitem{Auri89}  A. Aurilia, M. Palmer and E. Spallucci, Phys. Rev. D
\textbf{40}, 2511 (1989).

\bibitem{Land88}  See, e.g., L.D. Landau and E.M. Lifshitz ``The classical
theory of fields'', Fizmatlit, Moscow, 7th edition, (1988), Sec. 102.

\bibitem{Widr89}  L.M. Widrow, Phys. Rev. D \textbf{40}, 1002 (1989).

\bibitem{Auri91}  A. Aurilia, R. Balbinot and E. Spallucci, Phys. Lett. B
\textbf{262}, 222 (1991).

\bibitem{Balt91}  V. Baltar, J. Llamb\'ias and L. Masperi, Phys. Rev. D
\textbf{44}, 1214 (1991).

\bibitem{Ague95}  M.A. Ag\"uero-Granados, Phys. Lett. A \textbf{199}, 185
(1995).

\bibitem{Bere88}  V.A. Berezin, N.G. Kozimirov, V.A. Kuzmin and I.I.
Tkachev, Phys. Lett. B \textbf{212}, 415 (1988).

\bibitem{Auri90}  A. Aurilia and E. Spallucci, Phys. Lett. B \textbf{251},
39 (1990).

\bibitem{stat}  We want to mention that the stationary solution of Eq. (\ref
{bL3}) is time-independent in the interior coordinate frame; it might be
nonstationary in other coordinate systems.

\bibitem{Coll76}  P.A. Collins and R.W. Tucker, Nucl. Phys. B \textbf{112},
150 (1976).

\bibitem{Bere89}  V.B. Berestetskii, E.M. Lifshitz and L.P. Pitaevskii
\textquotedblleft Quatum electrodynamics\textquotedblright , Fizmatlit,
Moscow, 3rd edition, (1989), Sec. 45.

\bibitem{Kuch05}  Exponentially small penetration of the wave function into
the inside region was recently discussed in connection with reflection on
event horizon; see, e.g., M.Yu. Kuchiev and V.V. Flambaum, e-print
gr-qc/0502117.

\bibitem{Land01}  L.D. Landau and E.M. Lifshits, ``Quantum mechanics'',
Fizmatlit, Moscow, 5th edition, (2001), Sec. 46, 47.

\bibitem{Misn98}  C.W. Misner, K.S. Thorne and J.A. Wheeler,
``Gravitation'', Freeman and Company, New York, (1998), p. 783.

\bibitem{Cana00}  G. Canalizo and A. Stockton, ApJ \textbf{528,} 201 (2000).

\bibitem{Hipp96}  H. Hippelein, K. Meisenheimer and H.J. Ro\"ser, A\&A
\textbf{316}, 29 (1996).

\bibitem{Cheu02}  C.C. Cheung, ApJ \textbf{581,} L15 (2002).

\bibitem{Arp90a}  H. Arp and G. Burbidge, ApJL \textbf{353,} L1 (1990).

\bibitem{Arp94}  H. Arp, Proc. IAU Sym. No 168, Kluwer Academic Publishers,
Dordrecht, 1994, p. 401.

\bibitem{Chod85}  A. Chodos, A.I. Hauser and A. Kosteleck\'y, Phys. Lett. B
\textbf{150}, 431 (1985).

\bibitem{Moha98}  R.N. Mohapatra, Prog. Partcl. Nucl. Phys. \textbf{40}, 55
(1998).

\bibitem{Shiu02}  G. Shiu and I. Wasserman, Phys. Lett. B \textbf{541}, 6
(2002).

\bibitem{Bagl03}  J.S. Bagla, H.K. Jassal and T. Padmanabhan, Phys. Rev. D
\textbf{67}, 063504 (2003).

\bibitem{Davi04}  P.C.W. Davies, Int. J. Theor. Phys. \textbf{43}, 141
(2004).

\bibitem{Padm02}  T. Padmanabhan and T. Roy Choudhury, Phys. Rev. D
\textbf{66}, 081301 (2002).

\bibitem{Caus04}  M.B. Causse, astro-ph/0312206.

\bibitem{Lope04}  M. Lo\'pez and C.M. Gutie\'rrez, A\&A \textbf{421}, 407
(2004).

\bibitem{Aszt02}  S.J. Asztalos et al., ApJ \textbf{571}, L27 (2002).

\bibitem{Aszt04}  S.J. Asztalos et al., Phys. Rev. D \textbf{69}, 011101(R)
(2004).

\bibitem{Blou01}  B.D. Blout, E.J. Daw, M.P. Decowski, P.T.P. Ho, L.J.
Rosenberg and D.B. Yu, ApJ \textbf{546}, 825 (2001).

\bibitem{Kohl02}  M. K\"ohl, M.J. Davis, C.W. Gardiner, T.W. H\"ansch and T.
Esslinger, Phys. Rev. Lett. \textbf{88}, 080402 (2002).

\bibitem{Kolb94}  E.W. Kolb and I.I. Tkachev, Phys. Rev. D \textbf{49}, 5040
(1994).

\bibitem{Seid94}  E. Seidel and W.M. Suen, Phys. Lett. \textbf{72}, 2516
(1994).

\bibitem{PVLAS}   G. Cantatore et al. {\it ``First results on dichroism
induced by magnetic field in vacuum'', }talk presented at the National
Institute for Nuclear Physics, Trieste, Italy, 4 May, 2005;
http://www.ts.infn.it/physics/attivita\_scientifica\_2005/

\bibitem{Burb03a}  G. Burbidge, ApJ \textbf{586}, L111 (2003).

\bibitem{Whit00}  R.L. White et al., ApJS \textbf{126,} 133 (2000).

\bibitem{Beck01}  R.H. Becker et al., ApJS \textbf{135,} 227 (2001).

\bibitem{Sone77}  R.M. Soneira and P.J.E. Peebles, ApJ \textbf{211}, 1
(1977).

\bibitem{Arp97}  H. Arp, J. Astrophys. Astr. \textbf{18,} 393 (1997).

\bibitem{Arp98a}  H. Arp, ApJ \textbf{496,} 661 (1998).

\bibitem{Malo87a}  B.A. Malomed, Physica \textbf{24}D, 155 (1987).

\bibitem{Malo87b}  B.A. Malomed, Phys. Lett. A \textbf{123}, 459 (1987).
\end{thebibliography}
\end{document}